\documentclass[12pt,preprint]{aastex}
\slugcomment{Accepted for publication in The Astronomical Journal}  
\shorttitle{Stellar Kinematics Boxy Bulges}
\shortauthors{Chung \& Bureau}
\begin{document}
%
%
\title{Stellar Kinematics of Boxy Bulges:\\ Large-Scale Bars and Inner Disks}
\author{Aeree Chung\altaffilmark{1}} 
\affil{Department of Astronomy, Columbia University, 550~West
120th~Street, 1411 Pupin Hall, \\MC~5246, New York, NY~10027}
\email{archung@astro.columbia.edu}
\and 
\author{M.\ Bureau\altaffilmark{2}}
\affil{Columbia Astrophysics Laboratory, 550~West 120th~Street, 1027
 Pupin Hall, \\MC~5247, New York, NY~10027}
\email{bureau@astro.columbia.edu}
\altaffiltext{1}{Send offprint requests to Aeree Chung: archung@astro.columbia.edu}
\altaffiltext{2}{Hubble Fellow}
%
%
\begin{abstract} 
Long-slit stellar kinematic observations were obtained along the
major-axis of $30$ edge-on spiral galaxies, $24$ with a boxy or
peanut-shaped (B/PS) bulge and $6$ with other bulge types for
comparison. B/PS bulges are identified in at least $45\%$ of highly
inclined systems and a growing body of theoretical and observational
work suggests that they are the edge-on projection of thickened
bars. Profiles of the mean stellar velocity $V$, the velocity
dispersion $\sigma$, as well as the asymmetric ($h_3$) and symmetric
($h_4$) deviations from a pure Gaussian are presented for all
objects. Comparing those profiles with stellar kinematic bar
diagnostics developed from N-body simulations, we find bar signatures
in $24$ of our sample galaxies ($80\%$). Galaxies with a B/PS bulge
typically show a double-hump rotation curve with an intermediate dip
or plateau. They also frequently show a rather flat central velocity
dispersion profile accompanied by a secondary peak or plateau, and
numerous galaxies have a local central $\sigma$ minimum
($\gtrsim40\%$). The $h_3$ profiles display up to three slope
reversals. Most importantly, $h_3$ is normally correlated with $V$
over the presumed bar length, contrary to expectations from
axisymmetric disks. These characteristic bar signatures strengthen the
case for a close relationship between B/PS bulges and bars and leave
little room for other explanations of the bulges' shape. We also find
that $h_3$ is anti-correlated with $V$ in the very center of most
galaxies ($\gtrsim60\%$), indicating that these objects additionally
harbor cold and dense decoupled (quasi-)axisymmetric central stellar
disks, which may be related to the central light peaks. These central
disks coincide with previously identified star-forming ionized-gas
disks (nuclear spirals) in gas-rich systems, and we argue that they
formed out of gas accumulated by the bar at its center through
inflow. As suggested by N-body models, the asymmetry of the velocity
profile ($h_3$) appears to be a reliable tracer of asymmetries in
disks, allowing us to discriminate between axisymmetric and barred
disks seen in projection. B/PS bulges (and thus a large fraction of
all bulges) appear to be made-up mostly of disk material which has
acquired a large vertical extent through bar-driven vertical
instabilities. Their formation is thus probably dominated by secular
evolution processes rather than merging.
\end{abstract}
\keywords{galaxies: kinematics and dynamics --- galaxies: bulges ---
galaxies: spiral --- galaxies: structure --- galaxies: formation ---
galaxies: evolution}
%
%
\section{Introduction\label{sec:intro}}
The number of edge-on disk galaxies with a boxy or peanut-shaped
(B/PS) bulge is known to be significant. In fact, the fraction of
these objects has continuously increased as sample selection, data,
and classification criteria improved
\citep*[see][]{j86,sa87,s87,db88,ldp00a,ldp00b}. Recently,
\citet{ldp00a} classified $\approx1350$ edge-on disk galaxies and
found $45\%$ of B/PS bulges, with a fraction higher than $40\%$ for
all morphological types from S0 to Sd. This was the first study to
report such a large incidence of B/PS bulges in very late type
systems, but it is also the largest and most homogeneous. B/PS bulges
are therefore common, but they are only seen in highly inclined
systems ($i\gtrsim75^{\circ}$; e.g.\ \citealt*{sdb90}), indicating
that their shape is mainly related to the vertical distribution of
light.

Several scenarios have been suggested to explain the structure and
formation of B/PS bulges. They can be divided into two main
categories: axisymmetric accretion and non-axisymmetric bar buckling
models. \citet{bp85} and \citeauthor{r88} (\citeyear{r88}; but see
also \citealt{man85}) showed that it is possible to form axisymmetric
cylindrically rotating B/PS bulges from the accretion of one or (more
likely) many satellite galaxies onto a larger galaxy host. As pointed
out by \citet{bp85}, however, the particular alignment required puts
too many constraints on the orbits, angular momenta, and relative
masses of the galaxies, and this process appears unlikely in view of
the high incidence of B/PS bulges. Moreover, satellite galaxies are
not preferentially observed around galaxies with a B/PS bulge
\citep{s87}. Using N-body simulations, \citet{cs81} first suggested
that B/PS bulges could be due to the thickening of bars in the disks
of spiral galaxies. They showed that soon after a bar develops through
the usual bar instability, it buckles, thickens, and settles with a
larger velocity dispersion and thickness. When highly inclined, the
bar then appears peanut-shaped when seen side-on (i.e.\ perpendicular
to the bar), boxy-shaped when seen at intermediate angles, and almost
round when seen end-on (i.e.\ parallel to the bar). It is also largely
cylindrically rotating. \citeauthor{cs81}'s (\citeyear{cs81}) results
were later confirmed and improved upon by many studies
\citep[e.g.][]{cdfp90,rsjk91,am02} and they are now universally
accepted. The three-dimensional (3D) orbital structure of these bars
has just been investigated in a series of papers
(\citealt*{spa02a,spa02b,psa02,psa03}; but see also
\citealt{p84,pf91}), allowing a more detailed comparison of models
with high spatial resolution images.

There are several observational studies supporting the connection
between B/PS bulges and bars. Most important for this work,
\citet{km95} suggested that bars could be detected in edge-on galaxies
from their (projected) kinematic signature. In their pioneering study
of two spirals with a peanut-shaped bulge (\objectname{NGC~5746} and
\objectname{NGC~5965}), \citet{km95} found a characteristic
``figure-of-eight'' bar signature in their position-velocity diagrams
(PVDs; the projected density of material as a function of
line-of-sight velocity and projected position). Although the
interpretation of this signature was ultimately flawed, hydrodynamical
simulations by \citet{ab99} confirmed its relationship to bars and
showed how it can be used to identify and characterize edge-on bars
from gas kinematics. In a study of emission-line spectra for $10$
edge-on spirals with various types of bulges, \citet{mk99} later
confirmed the link between B/PS bulges and bars. \citeauthor{bf99}
(\citeyear{bf99}; see also \citealt{b98}) also presented ionized-gas
PVDs for $23$ edge-on galaxies, $17$ with a B/PS bulge and $6$ with
other types of bulges (mostly spheroidal). Using \citeauthor{ab99}'s
(\citeyear{ab99}) diagnostics, they found bar signatures in $14$ of
$17$ galaxies with a B/PS bulge ($\approx80\%$) and in none of the
galaxies with a more spheroidal bulge.

In spite of the aforementioned studies, our knowledge of the stellar
kinematics of B/PS bulges and their hosts remains limited. This
probably arises because the gas is a more practical tracer of the
barred potential, being both easier to observe and reacting more
strongly than the stars to asymmetries (shocks). Studying the stars
directly is nevertheless essential as B/PS bulges are stellar
structures and, when comparing to N-body simulations, stellar
kinematics largely bypasses issues such as the selection of an optimal
gaseous tracer (H$\alpha$, CO, \ion{H}{1}, etc), star formation, and
(for gas-poor objects) the response of the orbital structure to a
massive gas disk \citep[e.g.][]{bhsf98}.

Existing stellar kinematic studies of B/PS bulges are limited to a few
objects and have largely focused on the issue of cylindrical rotation,
using slits offset from the major-axis or perpendicular to it
\citep*[e.g.][]{bc77,ki82,r86,j87,s93a,swc93}. As noted above,
however, this is consistent with both axisymmetric and barred
configurations, and so does not discriminate between the two dominant
formation scenarios. We advocate here instead the new kinematic bar
diagnostics of \citet{ba04}, which directly probe the shape of the
potential \citep[but see also][]{bg94,km95}.

The importance of B/PS bulges goes beyond their sheer number and
interesting morphology, as they may hold vital clues for our
understanding of bulge formation. There is growing evidence against
significant merger growth in many bulges (e.g.\
\citealt*{k93,apb95,mch03,bgdp03}) and a corresponding emphasis on
secular evolution processes, largely bar-driven
\citep[e.g.][]{wfmmb95,es02}. Most theoretical models involve the
growth of a central mass through bar inflow and/or possibly recurring
bar destruction \citep*[e.g.][]{pn90,fb93,fb95,nsh96}, although the
efficiency of bar dissolution mechanisms remains uncertain
(\citealt{ss03} and references therein). Nevertheless, because of
their probable relationship to bars, B/PS bulges are likely to play a
central role in those scenarios, and a proper understanding of their
structure and dynamics is essential to constrain them.

Our primary goal in this work is thus to study the stellar kinematics
of a statistically significant number of galaxies with a B/PS bulge,
simultaneously and independently verifying their relationship to bars
and probing for embedded structures. We study \citeauthor{bf99}'s
(\citeyear{bf99}) sample of $23$ edge-on spirals, $17$ with a B/PS
bulge and $6$ with other bulge types, as well as an additional $7$
B/PS bulges which showed little (or confined) emission and were only
briefly discussed by the authors. The latter offer the best comparison
with N-body models, being largely gas and dust free.

In \S~\ref{sec:diagnostics}, the stellar kinematic bar diagnostics of
\citet{ba99} and \citet{ba04} are summarized. In \S~\ref{sec:obs}, we
describe the sample, observations, as well as the reduction and
analysis of the data. The stellar kinematics of the $30$ galaxies is
presented briefly in \S~\ref{sec:results}, while
\S~\ref{sec:discussion} discusses in greater detail the structure and
dynamics of B/PS bulges and how they fit into bar-driven secular
evolution scenarios. We summarize our results and conclude briefly in
\S~\ref{sec:conclusion}.
%
%
\section{Bar Diagnostics\label{sec:diagnostics}}
As B/PS bulges are only visible in (quasi-)edge-on galaxies, a
reliable morphological identification of bars is impossible. A plateau
in the light distribution along the major-axis has often been argued
to trace bars (e.g.\ \citealt{wh84,cc87,ldp00b}; but see also
\citealt{ba04}), but axisymmetric structures such as rings and lenses
can also account for them. In fact, it is possible to construct
axisymmetric B/PS bulges with simple two-integral distribution
functions \citep{r88}. We thus need bar identification methods for
edge-on disks relying on kinematics, truly probing the shape of the
galactic potential.

Adopting a generic barred galaxy mass model, \citet{ba99} used
periodic orbit families as building blocks to produce synthetic PVDs
of barred galaxies seen edge-on. They showed that because of the
non-homogeneous distribution of orbits in a barred disk,
characteristic gaps occur in the PVDs between the signatures of the
different orbit families. These can be used to kinematically identify
an edge-on bar and constrain its viewing angle and mass
distribution. Unfortunately, those models are too primitive to be
compared with either the observed gas or stellar kinematics. The gas
does not follow periodic orbits, being affected primarily by shocks
and dissipation. This led \citet{ab99} to develop more reliable gas
kinematic diagnostics using the same mass model but full
hydrodynamical simulations. These were successfully used by
\citet{bf99} in their ionized-gas study. Stars can also move on
regular (quasi-periodic), irregular, and chaotic orbits and must of
course form a self-consistent system, effectively smearing out the
signatures predicted by \citet{ba99}. More refined stellar kinematic
models are thus needed.

\citet{ba04} developed fully self-consistent stellar kinematic bar
diagnostics for edge-on disks using high-quality N-body simulations
similar to those of \citet{am02} and \citet{a03}. The simulations
include a luminous disk and a live dark halo, but no initial bulge
component. For various viewing angles, (large) inclinations, and bar
strengths, \citet{ba04} extracted the stellar kinematics along a slit
aligned with the photometric major-axis, contrasting the results with
those of an axisymmetric disk. Like most observers, they parametrized
the kinematics with a Gauss-Hermite series \citep[see,
e.g.,][]{mf93,g93}, using the same routine as used here (although no
deconvolution is required in the N-body case; see
\S~\ref{sec:obs}). The characteristic bar signatures observed are
numerous: i) a major-axis light profile ($I$) with a quasi-exponential
central peak and a plateau at moderate radii; ii) a ``double-hump''
rotation curve, that is a mean velocity ($V$) profile showing an
initial rise followed by a plateau or slight decrease and a second
rise to the flat part of the rotation curve; iii) a flat-top or
weakly-peaked velocity dispersion ($\sigma$) profile with a secondary
peak or plateau and, occasionally, a local central minimum (strong
bars only); iv) an $h_3$ profile {\em correlated} with $V$ over the
projected bar length; and v) an $h_4$ profile with central and
secondary minima. Throughout this paper, $h_3$ and $h_4$ represent
respectively the asymmetric (skewness) and symmetric (kurtosis)
deviations of the line-of-sight velocity distribution (LOSVD) from a
pure Gaussian, while $V$ and $\sigma$ have their usual meanings (see
\S~\ref{sec:obs}).

Those bar signatures are restricted to the projected bar length and
are strongest when the bar is seen end-on, gradually weakening as the
bar is viewed more side-on and they are stretched out
\citep[see][]{ba04}. In other words, the plateaus in the light profile
and the rotation curve, the peak in the dispersion, and the $h_3-V$
correlation are all stronger when the bar is seen end-on. This is
particularly useful as it is opposite from the morphological bias,
whereby the characteristic peanut shape is seen for side-on
projections and indiscriminate round shapes are seen end-on. As
expected, the strengths of all the bar signatures also correlate with
the strength of the bar itself (and the associated formation of an
inner ring). Furthermore, while one can probably build axisymmetric
configurations giving rise to the $I$, $V$, or $\sigma$ signatures
individually (e.g.\ double-disk structures), their radial extents are
correlated \citep{ba04} and the behavior of $h_3$ appears
characteristic of non-axisymmetric systems (although it is not
uniquely related to them). Indeed, the LOSVDs of (inclined)
axisymmetric systems will generally have a steep prograde wing and a
tail of lower velocity material, yielding an anti-correlation of $h_3$
and $V$. For decreasing surface brightness profiles, non-axisymmetric
motions thus appear required for $h_3$ and $V$ to correlate.

In an ideal situation, one would like to use both morphological and
kinematic information to study the intrinsic structure of a galaxy. In
this paper, however, we will mainly rely on the aforementioned stellar
kinematic diagnostics for $V$, $\sigma$, and $h_3$. Our $h_4$ profiles
generally have too low signal-to-noise ratios $S/N$ to be of much use,
and we defer discussion of the light profiles to a forthcoming pair of
papers where high quality $K$-band observations will be presented
\citep*{ababdf04,aab04}.
%
%
\section{Observations and Data Reduction\label{sec:obs}}
\subsection{Sample\label{sec:sample}}
We use the sample of \citeauthor{bf99} (\citeyear{bf99}; see also
\citealt{b98}), consisting of $30$ edge-on spiral galaxies selected
from the catalogs of \citet{j86}, \citet{s87}, and \citeauthor{sa87}
(\citeyear{sa87}; disk galaxies with a B/PS bulge), and from that of
\citeauthor*{kkp93} (\citeyear{kkp93}; galaxies with extreme axial
ratios). All objects are visible from the south
($\delta\lesssim15^\circ$) and have bulges larger than $0\farcm6$ in
diameter, to have enough spatial resolution to resolve the expected
bar signatures with moderate seeing. In addition, for quick imaging
with a small-field near-infrared (NIR) camera, we required
$D_{25}\lesssim7\arcmin$ (where $D_{25}$ is the diameter at the
$25$~mag~arcsec$^{-2}$ isophotal level in $B$).

General properties of the sample galaxies are summarized in
Table~\ref{tab:prop}. About three-quarters of the galaxies either have
probable companions or are part of a group or cluster, including a few
with alignments apparently suitable for accretion
\citep[e.g.][]{bp85}. However, on the whole, the sample is not biased
either against or for galaxies in a dense environment. Of the $30$
galaxies, $24$ were identified to have a B/PS bulge by \citet{j86},
\citet{s87}, and \citet{sa87}. The other $6$ galaxies were also
selected from those catalogs but show more varied morphologies, mostly
spheroidal. Those $6$ objects thus constitutes a ``control'' sample
for comparison. As we have already mentioned, however, N-body
simulations indicate that a (thick) bar seen end-on will appear round,
so we can expect some of the control sample galaxies to be barred as
well. Indeed, in their study of the gas kinematics, \citet{bf99} found
weak bar signatures in $2$ of the $6$ control galaxies
(\objectname{NGC~3957} and \objectname{NGC~4703}).

\placetable{tab:prop}

Since the sample has been put together from several catalogs, the
classification of the bulges' shapes is likely to be
inhomogeneous. Fortunately, however, the more recent and larger survey
of \citet{ldp00a} includes all of our sample galaxies. Although their
classification is somewhat subjective, relying on visual inspection of
contour plots from galaxy images (like the other authors),
\citet{ldp00a} did apply a consistent judgment to the entire sample so
their classification should at least be homogeneous. We compare in
Table~\ref{tab:class} the classification of \citet{ldp00a} with that
adopted by \cite{bf99} from the catalogs of \citet{j86}, \citet{s87},
and \citet{sa87}. A few galaxies absent from the latter catalogs were
also classified by \cite{bf99} themselves (\objectname{PGC~44931},
\objectname{IC~5096}, \objectname{NGC~4703}, and
\objectname{NGC~5084}). The classifications are consistent for $85\%$
of the galaxies. Three objects (\objectname{NGC~1596},
\objectname{IC~5096}, and \objectname{ESO~240-G011}) are classified as
boxy by \cite{bf99} but as ellipsoidal by \citet{ldp00a}, and one
object the opposite (\objectname{NGC~3957}). Bar signatures were
subsequently found by \citet{bf99} in both \objectname{IC~5096} and
\objectname{ESO~240-G011}.

\placetable{tab:class}

The uncertainties related to the morphological identification of bars in
edge-on systems through the presence of a B/PS bulge underscore the
necessity of kinematic studies. Since all galaxies with extended
ionized-gas emission were previously studied by \citet{bf99}, we will
somewhat focus here on the $7$ objects with little (or confined)
emission. Those are all dust-free early-type spirals and should
provide high quality stellar kinematics. The other galaxies are
nevertheless important to confirm or infirm the kinematic bar
detections of \citet{bf99} and to study the effects of gas.

Although the classification of the bulges' shape is robust between the
optical and NIR \citep[e.g.][]{ldp00a}, NIR images are essential to
study their detailed structure because of significant extinction in
most objects \citep[e.g.][]{ldp00b}. Clearly then, the observations
from this paper should be reexamined in view of the $K$-band images to
be presented in \citet{ababdf04} and \citet{aab04}. Preliminary
results have appeared in \citet{aabbdvp03}.
\subsection{Observations and Data Reduction\label{sec:reduction}}
Sets of long-slit spectroscopic observations were obtained along the
major-axis of each sample galaxy with the Double Beam Spectrograph of
the 2.3~m telescope at Siding Spring Observatory, Australia. The
observations were carried out between 1995 December and 1997 May for
$39$ nights in total. The red and blue arms of the spectrograph each
cover about $950$~\AA\ and were centered, respectively, on the
H$\alpha$ $\lambda6563$ line and the Mg~$b$ $\lambda5170$ triplet for
a redshift of $2000$~km~s$^{-1}$. Total exposure times ranged from
$12\,000$ to $21\,000$~s per object, split into $1500$~s
exposures. When no emission line was detected within $3000$~s, the red
arm of the spectrograph was recentered on the \ion{Ca}{2}
$\lambda8600$ triplet. $1752\times532$ SITE ST-D06A thinned CCDs were
used for all observations. All galaxies were observed with a
$1\farcs8\times400\arcsec$ slit. When a strong dust lane was present,
the slit was moved just above the major-axis. We note however that the
objects were acquired by visually aligning the nucleus and major-axis
on the slit using a slit-viewing video camera. As the system was
rather insensitive, the positioning of the slit was not very accurate,
especially for low surface brightness objects.

Ionized-gas kinematics extracted from the red arm have already been
presented and discussed by \citet{bf99} for all objects with extended
emission. Here we will focus on the blue arm data, best suited to
study the stellar kinematics using the absorption lines of the Mg~$b$
triplet, H$\beta$ $\lambda4861$, and Fe. The dispersion was
$\approx0.55$~\AA~pix$^{-1}$, yielding a $2$-pixel spectral resolution
of $64$~km~s$^{-1}$ ($\sigma_{\rm inst}=27$~km~s$^{-1}$; see
below). The spatial sampling was $0\farcs9$~pixel$^{-1}$ with typical
seeings of $1\farcs0-1\farcs5$.

All data were reduced using standard procedures in IRAF (Image
Reduction and Analysis Facility). The (two-dimensional) spectra were
first bias-subtracted using virtual overscan regions and bias frames,
dark-subtracted (when necessary), and flat-fielded using flattened
continuum lamp exposures. Each spectrum was then wavelength-calibrated
using bracketing arc lamp exposures and the spectra were
simultaneously rebinned to a logarithmic wavelength (linear velocity)
scale, corresponding to about $32$~km~s$^{-1}$~pixel$^{-1}$ over the
observed wavelength range ($4760-5710$~\AA). The spectra were then
corrected for vignetting along the slit using sky exposures. Every
exposure was corrected for Earth's motion and relevant groups of
exposures registered and combined. The resulting spectrum of each
object was sky-subtracted using source-free regions along the slit and
continuum normalized using a third-order spline with a small number of
pieces. These two-dimensional spectra, from both galaxies and template
stars, form the basis of the subsequent analysis aiming to extract the
stellar kinematics.

Because only $3$ template stars were observed systematically over the
entire observing program, we selected the Fourier Correlation Quotient
(FCQ) algorithm of \citet{b90} to derive the stellar kinematics of the
galaxies. This method is based on the deconvolution of the peak of the
template-galaxy correlation function with the peak of the template
autocorrelation function and is relatively insensitive to template
mismatch \citep[see, e.g.,][]{mf93}. Furthermore, being recovered
non-parametrically, the LOSVDs can be easily compared with those
obtained from N-body simulations. We subsequently fitted the LOSVDs
with a Gauss-Hermite series \citep{mf93,g93}, yielding the mean
stellar velocity $V$ and the velocity dispersion $\sigma$, as well as the
asymmetric ($h_3$) and symmetric ($h_4$) deviations from a pure
Gaussian. This is essential given the complex shapes expected from the
profiles \citep[see][]{ba04}. We nevertheless fit $V$ and $\sigma$
independently of $h_3$ and $h_4$ so that our results are directly
comparable with single Gaussian fits. We use the deconvolution and
fitting algorithms as originally implemented in the XSAURON software
package \citep[see][]{betal01,zetal02}.

The resulting stellar kinematics for the $30$ sample galaxies are
presented in Figure~\ref{fig:kin_noem} (B/PS bulges with no or
confined emission), Figure~\ref{fig:kin_em} (B/PS bulges with extended
emission), and Figure~\ref{fig:kin_control} (control sample; all with
extended emission). The latter two figures can be directly compared to
those in \citet{bf99} for the gas kinematics. From top to bottom, each
panel shows an optical image of the galaxy (with isophotal contours)
and registered $V$, $\sigma$, $h_3$, and $h_4$ profiles along the
major-axis. The spatial scale varies from galaxy to galaxy but the
kinematic limits are the same. The data were folded under the
assumption of bisymmetry and the errors reported are half the
difference between the approaching and receding sides of each
galaxy. When the datum from one side is missing or corrupted (e.g.\
due to failure of the deconvolution algorithm or a foreground star),
the datum is mirrored and denoted by a cross without an error bar. To
ensure a homogeneous and minimal $S/N$ in the outer parts of the
galaxies, the data were binned until $S/N\ge5$ per spatial and
spectral element for $V$ and $\sigma$ and $S/N\ge10$ for $h_3$ and
$h_4$, explaining the different bins for those quantities. The final
$S/N$ is thus at least $5\times\sqrt{2}\approx7$ for $V$ and $\sigma$
and $10\times\sqrt{2}\approx14$ for $h_3$ and $h_4$.

\placefigure{fig:kin_noem}
\placefigure{fig:kin_em}
\placefigure{fig:kin_control}
\subsection{Data Quality Tests\label{sec:quality}}
We tested our results for various sources of errors and uncertainties:
strong sky emission lines, template mismatch, erroneous continuum
normalization, variable spectral focus, and strong galactic emission
lines.

A number of night sky emission lines from natural (e.g. \ion{N}{1},
\ion{O}{1}) and man-made (e.g.\ \ion{Hg}{1}, \ion{Na}{1}) sources are
present in the observed wavelength range \citep[see][]{mf00}. All were
successfully removed by our sky subtraction algorithm except for the
strongest \ion{O}{1} telluric emission line at $\lambda5577$, where
the resulting artifacts are sufficient to affect our measurements,
especially for faint galaxies. The spectra were thus truncated
shortward of $\lambda5577$, shortening the spectra by about
$150$~\AA. This barely affects the quality of our measurements as
there are no important galactic absorption lines longward of
$\lambda5577$ in our spectral range.

As gas kinematics was the primary goal of our observations, only two
different types of stars were observed: K1~III
(\objectname{CPD~-43$\degr$2527} and \objectname{HD~176047}) and G5~IV
(\objectname{HD~101266}). This seriously limits our ability to
quantify the effects of template mismatch, but the stellar kinematics
of all the objects was nevertheless derived using both types of star
and the results compared. As expected, no qualitative or significant
quantitative differences were found, except for $h_4$ which is known
to be particularly sensitive to mismatch \citep[e.g.][]{mf93}. Our
$h_4$ profiles are in any case noisy and will not be discussed in any
depth. The results presented throughout this paper were thus derived
using a K1~III star only, and the effects of template mismatch will
not be discussed further.

We also tested for the effects of imperfect continuum normalization by
comparing the stellar kinematics obtained from good and bad
normalizations of both galaxies and template stars. We used a linear
fit to the continuum for bad normalizations and a cubic spline of
order $8$ for good normalizations ($\approx100$~\AA ~per order; our
default value). We found that only a bad continuum normalization of
the template stars affected our results. But since template star
spectra have high $S/N$ and are few, obtaining a high quality
continuum subtraction is both fast and easy. Continuum normalization
is therefore not an issue for our data.

Although the instrumental setup was the same for all observing runs, a
number of parameters such as the spectral focus (spectral resolution)
were set manually and could have changed slightly from run to run. We
checked for any effect by comparing the kinematics of galaxies
observed over more than one run, and by deriving the kinematics of a
few objects using template stars observed over different runs. No
significant differences were found. A more worrying issue concerns the
spectral resolution near the ends of the slit, as it was not always
possible to obtain a uniform spectral focus over the entire slit
length. In certain runs, judging from arc lamp exposures, the spectral
resolution degraded significantly and the spectral
point-spread-function (PSF) became non-Gaussian at the slit
ends. Since template stars were only observed in the central parts of
the slit, this could result in artificially high velocity dispersions
and biased $h_3$ and $h_4$ measurements at large galactic radii. While
we have no direct way of quantifying these effects (except for
characterizing the widening of the arc lines themselves in arc lamp
exposures), only a few galaxies have stellar measurements extending
over a substantial portion of the slit, with $2$ galaxies just
reaching $50\%$ (\objectname{NGC~4710} and
\objectname{NGC~5746}). Those do not show any peculiarities or
systematic behavior at large radii, so we believe our measurements to
be reliable over the radial range probed in all objects. In any case,
our analysis will focus on the central regions of the galaxies only
(within two bulge diameters), well within the properly focused region.

Lastly, we tested for possible effects due to the presence of galactic
emission lines, which were not subtracted prior to deconvolution, and
can not be masked when using FCQ. The emission line of [\ion{O}{3}]
$\lambda5007$ is detected in most objects and that of H$\beta$
(superposed on the corresponding absorption line) in about half,
although both are rather weak in most cases. The [\ion{N}{1}]
$\lambda5200$ doublet is not detected in any object, allowing us to
compare the stellar kinematics obtained from a narrow range of
wavelengths around the Mg~$b$ triplet (unaffected by emission lines)
to that derived from the entire spectra. Using a short wavelength
range is not advised with FCQ but does not appear problematic
here. Using the same spatial bins, no significant difference is found
for the odd moments of the velocity profiles ($V$ and $h_3$), except
for increased noise in the shorter spectra as expected (and the
failure of the algorithm for \objectname{ESO~151-G004} and
\objectname{ESO~597-G036}, two of the three faintest
galaxies). However, as shown in Figure~\ref{fig:sigma_test}, the
$\sigma$ profiles derived from the shorter unaffected spectra are
systematically higher for about one-third of the objects (by typically
$20$~km~s$^{-1}$ but up to $100$~km~s$^{-1}$). This occurs over a
limited radial range, typically encompassing the rapidly rising part
of the rotation curve and never extending past the first velocity
plateau (see \S~\ref{sec:results}). It could indicate that the
kinematics of those objects is indeed affected by strong emission,
since \citet{bf99} identified numerous bright, rapidly rotating
ionized-gas disks (nuclear spirals) in those regions. However, there
is no direct relationship between the strength of the central emission
and the differences in velocity dispersions, suggesting a more complex
explanation. As we will argue later (see \S~\ref{sec:evolution}), it
is likely that the ionized-gas disks are associated with star
formation and a dynamically cold, young population of stars. These
would have little effect on the Mg~$b$ triplet but would strongly
influence the H$\beta$ line, biasing any measurement including the
latter to lower values (as observed). This also suggests that a single
late-type stellar template is not a sufficiently accurate
representation of the effective stellar type of many sample galaxies
over the entire radial range observed (although it usually does a fine
job in the outer parts). Fortunately, however, the qualitative
behavior of the profiles (on which our analysis relies) is not
substantially affected.

\placefigure{fig:sigma_test}
\subsection{Comparison with Published Data\label{sec:published_data}}
Comparisons of our data with published stellar and gas kinematics are
shown in Figures~\ref{fig:comp_stellar} and \ref{fig:comp_gas},
respectively. We found only $5$ galaxies with published stellar $V$
and/or $\sigma$ profiles (whether in tabular or figure formats):
\objectname{NGC~1381} \citep{ds83,dzlcc95}, \objectname{NGC~2310}
\citep{lzbr94}, \objectname{NGC~128} \citep{ds83,dzlcc95},
\objectname{NGC~5084} \citep{ds83}, and \objectname{NGC~5746}
\citep{brbh93}. See also \citet{bc77} and \citet{ea97} for
\objectname{NGC~128}. $9$ galaxies have available ionized-gas rotation
curves, all of them measured from H$\alpha$: \objectname{IC~4767}
\citep*{rwf88,abmbcs95}, \objectname{NGC~1886},
\objectname{NGC~2788A}, \objectname{IC~2531},
\objectname{ESO~443-G042}, \objectname{ESO~240-G011} \citep*{mfb92},
\objectname{NGC~4469}, \objectname{NGC~4710} \citep*{rwk99}, and
\objectname{ESO~597-G036} \citep{nsomt00}. We note that \citet{bf99}
did not extract rotation curves from their data, relying instead on
the entire PVD diagrams. Although the identification of the
approaching and receding sides in the literature is not always
consistent with ours, we use matching sides here for comparison.

\placefigure{fig:comp_stellar}
\placefigure{fig:comp_gas}

For the stellar kinematics, our data generally have better spatial
sampling and systematically extend to larger radii than published data
(see Fig.~\ref{fig:comp_stellar}). \objectname{NGC~1381} and
\objectname{NGC~128} are best suited for comparison since they have
recent measurements also derived with FCQ \citep{dzlcc95}. The
agreement is excellent, except for \objectname{NGC~1381} at large
radii where our $\sigma$ measurements are systematically
lower. However, we believe that this is due to \citeauthor{dzlcc95}'s
(\citeyear{dzlcc95}) lower spectral resolution ($\sigma_{\rm
inst}\gtrsim43$~km~s$^{-1}$ rather than $27$~km~s$^{-1}$ here) and we
favor our results. The discrepancies in $V$ for \objectname{NGC~128}
at large radii are not serious since the galaxy is disturbed at large
radii and our data are folded. \citeauthor{brbh93}'s
(\citeyear{brbh93}) measurements of \objectname{NGC~5746}, obtained by
cross-correlation \citep{td79}, are also in good agreement, except for
a few arcseconds at positive $V$ (one-sided feature). In the central
$15\arcsec$, $\sigma$ is in agreement with the values derived from
Mg~$b$ only. \citeauthor{ds83}'s (\citeyear{ds83}) data for
\objectname{NGC~1381}, \objectname{NGC~128}, and \objectname{NGC~5084}
only marginally agree with our, but they are of much lower quality,
have much lower spectral resolution ($\sigma_{\rm
inst}=140$~km~s$^{-1}$), and were taken through a much wider slit. Our
data should thus be considered superior. The only true problem lies
with \objectname{NGC~2310}, where our measurements strongly disagree
with literature data, also derived with FCQ but limited to the inner
$5\arcsec$. We have no convincing explanation for the differences but
offer a few hypotheses. Very bad seeing could explain our $V$ and
$\sigma$ measurements, systematically lower than those of
\citet{lzbr94}, but the effect seems too large. Another possibility is
that our slit was positioned slightly above (or below) the
major-axis. As discussed above, positioning the slit with precision
was difficult, but unless the vertical gradients of the mean velocity
and velocity dispersion are very steep, the magnitude of the effect is
unlikely to be sufficient. If \objectname{NGC~2310} is indeed barred
(see \S~\ref{sec:discussion}), these gradients are expected to be
shallow \citep[see, e.g.,][]{cdfp90}. Another possible explanation is
that the error bars from \citet{lzbr94}, which are not detailed but
are of comparable sizes to those of the other galaxies studied, are
underestimated. Indeed, \objectname{NGC~2310} was observed with one of
the shortest integration times ($2700$~s) and with the worst spectral
resolution ($\sigma_{\rm inst}\gtrsim142$~km~s$^{-1}$) of their entire
sample, and the dispersions could be systematically overestimated.

The comparison with published gas kinematics is presented for
reference only. The gas and stellar kinematics can not be easily
compared, so any discrepancy between ours and published data can not
be straightforwardly assigned to technical or reduction/analysis
problems. In particular, the gaseous tracers used in
Figure~\ref{fig:comp_gas} most likely lie in a rotationally supported
disk with small velocity dispersion, while stars can have significant
pressure support and correspondingly reduced rotation, especially in
the central parts. Furthermore, the gas can be strongly affected by
shocks, inflow, etc \citep[see][]{ab99}. All these effects are indeed
observed in Figure~\ref{fig:comp_gas}, where the agreement between our
stellar measurements and published gas kinematics is always good in
the outer parts but half the galaxies are discrepant in the central
regions. In the galaxies where the effect is strongest
(\objectname{IC~4767}, \objectname{NGC~4469}, \objectname{NGC~4710},
and \objectname{NGC~5746}), large B/PS bulges are present and
\citet{bf99} identified fast-rotating central gas components (nuclear
spirals), apparently associated with $x_2$-like streamlines
\citep{a92a,a92b}.
%
%
\section{Results\label{sec:results}}
\subsection{Rotation curves and asymmetries\label{sec:odd}}
A look at Figures~\ref{fig:kin_noem}--\ref{fig:kin_control} quickly
reveals that most galaxies with a B/PS bulges display a
``double-hump'' rotation curve, as expected from a barred disk (see
\S~\ref{sec:diagnostics}; \citealt{ba04}). The most extreme examples
are \objectname{NGC~6771}, \objectname{NGC~128}, and
\objectname{NGC~5746}, where the $V$ profile decreases significantly
after a steep initial rise, before rising again slowly to the flat
part of the rotation curve (the so-called turnover radius). Most
galaxies, however, show a rather flat velocity plateau at intermediate
radii (e.g.\ \objectname{NGC~1596}, \objectname{IC~4767}, and
\objectname{PGC~44931}), while a few merely show a change of slope in
the rising part of the rotation curve (e.g.\ \objectname{NGC~4710},
\objectname{NGC~4469}, and \objectname{IC~5096}). Only $3$ of the $24$
galaxies with a B/PS bulge show no evidence of a double-hump $V$
profile: \objectname{NGC~3390}, \objectname{ESO~443-G042}, and
\objectname{ESO~240-G011}. One could argue that this is also the case
for \objectname{NGC~1886} and \objectname{IC~2531}, but the latter
(like \objectname{ESO~443-G042}) is heavily obscured by dust and the
extracted kinematics is unreliable. Most galaxies of the control
sample ($4/6$) show a featureless rotation curve smoothly rising in
the inner parts, as expected for an axisymmetric object
\citep[e.g.][]{ba04}. \objectname{NGC~3957} and \objectname{NGC~4703},
however, do show a double-hump feature.

Since double-hump velocity profiles can equally be explained by
carefully-constructed axisymmetric models (e.g.\ double-disk
structures), the $h_3$ profiles are crucial to argue for triaxiality
\citep{ba04}. The quality of the $h_3$ profiles varies enormously
across the sample, from excellent (e.g.\ \objectname{NGC~1381},
\objectname{NGC~5746}, and \objectname{NGC~6722}) to almost useless
(e.g.\ \objectname{NGC~1886}, \objectname{ESO~443-G042}, and
\objectname{NGC~2788A}). For essentially all B/PS bulges with an
acceptable $h_3$ profile (arguably $20/21$ galaxies), $h_3$ is {\em
correlated} with $V$ over a large radial range in the central parts,
as expected from a projected barred disk (but contrary to expectations
for an axisymmetric one; see \S~\ref{sec:diagnostics} and
\citealt{ba04}). For most of those objects ($17/20$ galaxies),
however, this correlation is not valid over the entire bulge region,
but only roughly within the flat part of the rotation curve at
moderate radii (the first $V$ plateau). Most objects show an $h_3-V$
anti-correlation over a short radial range in the very center,
corresponding approximately to the rapidly rising part of the rotation
curve. The best examples of this are \objectname{NGC~1381},
\objectname{NGC~3203}, \objectname{NGC~128}, and
\objectname{NGC~5746}, while some of the weakest trends are seen in
\objectname{IC~4767}, \objectname{ESO~185-G053}, and
\objectname{NGC~4469}. Only $3$ of $21$ galaxies do not show this
central anti-correlated component, and the $h_3$ profile of
\objectname{ESO~597-G036} is anti-correlated over its entire radial
range. We note also that outside the range where $h_3$ is correlated
with $V$ (i.e.\ from the second rise in the rotation curve), $h_3$ and
$V$ are generally anti-correlated again. Two $h_3$ slope reversals are
therefore the norm, and a third one is often seen in the outer disk
(e.g.\ \objectname{NGC~1596}, \objectname{NGC~3203},
\objectname{NGC~5746}, and \objectname{NGC~6722}). Surprisingly, only
$3$ of the $6$ galaxies in the control sample behave as expected for
axisymmetric objects (\objectname{NGC~1032}, \objectname{NGC~5084},
and \objectname{IC~5176}). \objectname{NGC~3957} and
\objectname{NGC~4703} are again bar-like, while \objectname{NGC~7123}
shows a region where $h_3$ may be correlated with $V$ (although the
profile is unreliable, most likely due to dust).

The $V$ and $h_3$ trends are perhaps better illustrated in
Figures~\ref{fig:v/sigma} and \ref{fig:h3}, where the dimensionless
parameter $V/\sigma$ is shown as a function of the projected radius
and $h_3$ is shown as a function of $V/\sigma$, respectively. The rise
of $V/\sigma$ is nearly linear with radius for all galaxies, although
it flattens out in the outer parts for some. $h_3$ and $V/\sigma$ are
generally anti-correlated at small and large $V/\sigma$ values (small
and large radii), while they are correlated at moderate $V/\sigma$
(moderate radii), where the bar is thought to dominate (see
\S~\ref{sec:discussion}). Figure~\ref{fig:schem_diagram} provides a
schematized view of the $V$, $\sigma$, and $h_3$ profiles summarizing
the trends observed in B/PS bulges.

\placefigure{fig:v/sigma}
\placefigure{fig:h3}

We note that we have verified that the $h_3-V$ (anti-)correlations can
not be introduced by folding the data. If this was the case, the error
bars as defined would clearly indicate it (e.g.\ they would always
cross the $h_3=0$ line) and we would consider the (anti-)correlations
insignificant. The same behavior is observed in the non-folded
profiles, only at lower $S/N$.
\subsection{Dispersions and symmetries\label{sec:even}}
There is a great variety in the observed velocity dispersion
profiles. A few galaxies possess a flat central $\sigma$ peak with a
broad shoulder and a small secondary peak (e.g.\ \objectname{IC~4767}
and \objectname{ESO~185-G053}). Most galaxies, however, have a sharper
central peak with a gently decreasing shoulder, sometimes showing
hints of a secondary maximum. Typical examples include
\objectname{NGC~1381}, \objectname{NGC~3203}, and
\objectname{NGC~4469}. N-body simulations predict both types of
behavior, the former for stronger and the latter for weaker bars
\citep[see][]{am02,ba04}. Yet other galaxies only show a dominant,
broad central peak with a clear secondary peak (e.g.\
\objectname{NGC~128}, \objectname{NGC~5746}, and
\objectname{NGC~6722}). The secondary dispersion maximum is usually
strongest in the galaxies with a strong double-hump velocity profile
(a presumably strong bar).

Many $\sigma$ profiles also show a central depression, that is a more
or less extended local central minimum. The best examples are
\objectname{NGC~5746}, \objectname{NGC~4469}, and
\objectname{NGC~4710}, where the depression extends over several
arcseconds, but many galaxies show a central minimum in the central
$1-2$ pixels only (e.g.\ \objectname{NGC~128}, \objectname{NGC~2788A},
and \objectname{NGC~5084}). Similar local $\sigma$ minima were
observed in the N-body simulations of \citet{ba04} but were generally
restricted to the strongest bars, so we will discuss them at more
length in \S~\ref{sec:disks}. In total, $10/24$ galaxies with a B/PS
bulge and $3/6$ galaxies from the control sample have a clear local
central $\sigma$ minimum, while a few more have marginal
signatures. This result is not affected by the spectral range used
(whole spectra or Mg~$b$ region only; see Fig.~\ref{fig:sigma_test})
and is particularly surprising since local central $\sigma$ minima are
generally thought to be rare, although it may be due to dust in a few
cases (e.g.\ \objectname{NGC~2788A} and \objectname{NGC~7123}). In the
velocity dispersion profile compilation of \citet{b93}, only one clear
example is present out of $12$ intermediate-type spirals. On the other
hand, a central $\sigma$ minimum is visible in $7$ of the $18$ S0s
studied by \citet{f97}, entirely consistent with our result. There,
the minimum is only apparent along the minor-axis for $4$ objects,
suggesting that we would detect an even larger fraction of galaxies
with a local central $\sigma$ minimum if we had minor-axis
profiles. Such a situation can easily arise if the kinematically cold
component is more flattened than the spheroid
\citep[e.g.][]{zetal02,eetal04}. Recently, \citet{egcflpw01} report a
central $\sigma$ minimum in $3$ of $4$ intermediate-type spirals
studied, suggesting that the fraction increases as the quality (and
most importantly the spatial resolution) of the data increases.

The $h_4$ profiles of the sample galaxies generally have low $S/N$ and
their measurement is somewhat affected by emission and template
mismatch. We therefore refrain from a detailed discussion of their
properties. Let us simply note that the inner parts of the profiles
are generally correlated with $\sigma$ (e.g.\ \objectname{NGC~128} and
\objectname{NGC~5746}), but that there are a few exceptions (e.g.\
\objectname{NGC~1381} and \objectname{ESO~240-G011}). No correlation
with the bulge shape or other kinematic quantities is apparent.
%
%
\section{Discussion\label{sec:discussion}}
\subsection{B/PS Bulges and Bars\label{sec:bars}}
Our observations clearly show that most galaxies with a B/PS bulge
display the stellar kinematic signatures expected from a barred disk
seen edge-on (see \S~\ref{sec:diagnostics}; \citealt{ba04}). Indeed,
as sketched in Figure~\ref{fig:schem_diagram}, most galaxies in
Figures~\ref{fig:kin_noem} and \ref{fig:kin_em} show i) a double-hump
rotation curve, ii) a rather flat or slightly peaked velocity
dispersion profile with a broad shoulder (or secondary maximum) and,
occasionally, a local central minimum, and iii) an $h_3$ profile
correlated with $V$ over the expected bar length (the first hump of
the rotation curve). These results are summarized in the last column
of Table~\ref{tab:class}, where the sample galaxies are labeled as
``bar'', ``bar?'', or ``no bar'' if they show at least $2$, only $1$,
or none of the above $3$ kinematic bar signatures.

\placefigure{fig:schem_diagram}

Although it holds true for all B/PS bulges, the link with bars is
clearest for the sample of galaxies with no (or confined) emission
(Fig.~\ref{fig:kin_noem}), as well as for the earliest type galaxies
in the second B/PS bulge sample (the S0s of Fig.~\ref{fig:kin_em}),
where the emission is very weak outside of the central ionized-gas
disk (the nuclear spiral; see \citealt{bf99}). Except for
\objectname{ESO~597-G036}, which is strongly interacting, those
galaxies systematically show all $3$ kinematic bar
signatures. Furthermore, the bulges' morphology does seem to correlate
with the shape of the kinematic profiles (and thus the bar viewing
angle) as expected (see \S~\ref{sec:diagnostics}; \citealt{ba04}). The
strongest peanut-shaped bulges (e.g.\ \objectname{IC~4767} and
\objectname{ESO~185-G053}) show an extended first velocity hump
(relative to the bulge size), a rather flat $\sigma$ profile, and
small $h_3$ amplitudes; rounder bulges (e.g.\ \objectname{NGC~1381}
and \objectname{NGC~1596}) have comparatively shorter velocity
plateaus, more peaked $\sigma$ profiles, and strong $h_3$
signatures. Furthermore, the galaxies with the strongest velocity
plateaus (or dips) generally show the best evidence for a secondary
dispersion peak (e.g.\ \objectname{NGC~128} and
\objectname{NGC~6771}).

Those morphology--viewing angle correlations are not in prefect
agreement with the simulations' predictions, however, especially for
galaxies with extended emission. Many peanut-shaped bulges show strong
$V$ double-humps, peaked $\sigma$ profiles, and clear $h_3-V$
correlations, in accordance with the presence of a bar but contrary to
the viewing angle (or morphology) expectations \citep{ba04}. This is
best illustrated by \objectname{NGC~128}, \objectname{NGC~5746}, and
\objectname{NGC~6771}, but see also \objectname{NGC~6722},
\objectname{PGC~44931}, and \objectname{NGC~2788A}. This suggests that
the relationship between bulge morphology and bar viewing angle is not
as straightforward as the N-body simulations suggest
\citep[e.g.][]{cdfp90,rsjk91}, or that the pure N-body kinematics is
too idealized, particularly for gas-rich galaxies. The structure and
dynamics of (thick) bars may be strongly influenced by the presence of
a gas disk, as has been advocated by \citet{bhsf98}. While they argued
that the bar and buckling instabilities (and thus the boxiness of the
bulges) are weakened by gas, which does not seem to be the case here
(there are plenty of gas-rich but strong B/PS bulges), it would be
interesting to see if the kinematic bar signatures are strongly
affected.

Lessening the discrepancy between simulations and observations for the
later type galaxies is the fact that the kinematic bar signatures are
modified for non edge-on systems, as the line-of-sight integration
through the outer parts of the disk decreases \citep[see][]{ba04}. For
those cases, the $V$ and $h_3$ signatures become more extreme and the
secondary $\sigma$ peak increases, particularly for side-on
bar orientations. As the gas-poor early-type galaxies of the sample are
perfectly edge-on, while the gas-rich late-type galaxies are generally
less inclined (to avoid the dust lane), the latter are expected to
have stronger signatures (in a relative sense). There does not seem to
be an obvious correlation between inclination and bar signature
strength, but the effects of dust are hard to quantify and some of the
strongest bar signatures are indeed seen in non edge-on systems. The
effect is thus likely present, but it is unclear whether it can
account for the very strong bar signatures observed in objects such as
\objectname{NGC~5746} and \objectname{NGC~6722}.

The complexity of the kinematic profiles can also be artificially
increased by the presence of dust along the line-of-sight. The
kinematics of a few galaxies with perfectly edge-on and strong dust
lanes is probably unreliable (e.g.\ \objectname{IC~2531} and
\objectname{ESO~443-G042}) and other galaxies are likely to be
affected to various degrees (e.g.\ \objectname{NGC~2788A} and
\objectname{IC~4937}). The apparent shape of the bulges may also be
modified, although their classification as B/PS bulges is probably
unaffected \citep{ldp00b}. To alleviate the extinction problem, we
have acquired $K$-band imaging of all sample galaxies. A separate
study of their morphology will appear in \citet{ababdf04}
and \citet{aab04}, and the gas and stellar kinematics of the sample
shall be reexamined in light of those results. $K$-band images will be
especially useful to constrain the scales over which the kinematic bar
signatures occur. Currently, the extent of the features in $V$,
$\sigma$, and $h_3$ can be compared among themselves, but they can
only be roughly related to the relevant photometric scales of the
bulge and disk.

To summarize, despite kinematic profiles in appearance more complex
than those predicted by N-body simulations, our observations
convincingly show that most galaxies with a B/PS bulge have a barred
disk. This is because most galaxies possess the $V$, $\sigma$, and
particularly the $h_3$ signatures expected from an edge-on bar
(\S~\ref{sec:diagnostics} and \citealt{ba04}). This in turn strongly
supports the suggestion that B/PS bulges are truly thick bars seen
edge-on, although we have not probed the galaxies' potential out of
the disk plane. Kinematics at large galactic heights is required for
the latter, and we have acquired data for a few objects (Zamojski et
al.\ 2004, in preparation).

As noted in \S~\ref{sec:intro}, some off-plane kinematic observations
of B/PS bulges do exist in the literature
\citep[e.g.][]{bc77,ki82,r86,j87,s93a,swc93}, but those were mostly
concerned with the presence of cylindrical rotation rather than the
current bar signatures. Cylindrical rotation is a generic prediction
of (thick) bar models \citep[e.g.][]{cdfp90}, but it is also
consistent with axisymmetry \citep{r88}.

Interestingly, the galaxies in the control sample do not
systematically show the behavior expected for a simple axisymmetric
disk: i) a smoothly rising rotation curve, ii) a monotonically
decreasing velocity dispersion profile, and iii) an $h_3$ profile
anti-correlated with $V$ \citep[see][]{ba04}. Some of those galaxies
may well have substructures, but the problem most likely arises from
the heterogeneous nature of the control sample. Indeed, while there
are a few catalogs of galaxies with a B/PS bulge (see
\S~\ref{sec:sample}), there is no such thing for edge-on round
bulges. The control sample was thus selected from the catalog of high
axial ratio galaxies by \citet{kkp93} and from galaxies apparently
misclassified in B/PS bulge catalogs. This is less than satisfying,
and \objectname{NGC~3957} and \objectname{NGC~4703} clearly represent
transition objects (weak B/PS bulges and thus probably weak
bars). Those $2$ objects have the ``boxiest'' bulges of the control
sample and show characteristic kinematic bar signatures. \citet{bf99}
already pointed out that they have weak bar signatures in their
ionized-gas kinematics. To confirm that most non-B/PS bulges are also
non-barred (but not all, since some are expected to be if the bar is
seen end-on), a study similar to ours but focusing instead on round
bulges would be highly valuable. A larger and better defined control
sample is thus essential to strengthen our results.
\subsection{Inner Disks\label{sec:disks}}
As mentioned above, at least $10$ of the $24$ galaxies with a B/PS
bulge ($\gtrsim40\%$) show a decrease of $\sigma$ in the very
center. In a few galaxies, this local minimum extends over several
arcseconds and can reach $15-20\%$ of the observed peak (e.g.\
\objectname{NGC~5746}, \objectname{NGC~4469}, and
\objectname{NGC~4710}), although it is generally much smaller in both
size and intensity (e.g.\ \objectname{NGC~128} and
\objectname{NGC~2788A}). $8$ of those galaxies also show a
characteristic double-hump velocity profile. The standard explanation
for both features is the presence of a cold central stellar disk
originating from gas inflow \citep[e.g.][]{egcflpw01}. However,
\citet{am02} noted the existence of central $\sigma$ minima in purely
dissipationless N-body simulations. \citet{ba04} later showed that
they can arise naturally from the orbital structure of (strongly)
barred disks (depending on the axial ratio radial dependence of the
$x_1$ orbits). We have so far argued for the latter possibility in
this paper because of the $h_3-V$ correlation at moderate radii (i.e.\
over the expected bar length), which is characteristic of triaxiality
and bypasses the need for gas \citep{ba04}.

However, the observations do show a strong anti-correlation of $h_3$
and $V$ in the inner parts of most sample galaxies (including all
those with a central $\sigma$ minimum and a reliable $h_3$ profile;
e.g.\ \objectname{NGC~5746}, \objectname{NGC~4469}, and
\objectname{NGC~128}). This behavior is not observed in any of the
N-body simulations of \citet{ba04} and suggests that not only are the
inner parts of the galaxies cold (i.e.\ rapidly rotating), but also
that they are very close to axisymmetric. In those objects, the region
where $h_3$ and $V$ anti-correlate appears to have largely decoupled
from the rest of the galaxy (and the bar) and circularized, forming a
dense and (quasi-)axisymmetric central stellar disk. This is supported
by the sharp transition between the $h_3-V$ anti-correlation and
correlation regions. Furthermore, (quasi-)axisymmetric central stellar
disks naturally explain why prominent double-hump $V$ profiles and
central $\sigma$ minima are observed in peanut-shaped bulges as well
as boxy-shaped ones, contrary to viewing angle expectations. Once
circularized, the kinematic signatures of the inner regions will be
largely independent of viewing angle (but not exactly, because of the
line-of-sight integration through the main disk, which remains
non-axisymmetric). 

As expected, the extent of the central $h_3-V$ anti-correlation is
roughly equal to that of the rapidly rising part of the rotation
curve. If present, the extent of the central $\sigma$ minimum is also
equal or smaller (see Fig.~\ref{fig:schem_diagram};
\objectname{NGC~5746}, \objectname{NGC~4469}, and
\objectname{NGC~4710}). Of the $3$ galaxies in the control sample which
show a central $\sigma$ depression, \objectname{NGC~7123} is probably
affected by dust, but \objectname{NGC~5084} and \objectname{IC~5176}
appear as regular objects with large-scale axisymmetric disks (and no
$h_3-V$ correlation).

As our kinematic profiles are luminosity-weighted, the central disks
could in principle be fairly bright but only marginally colder than
their surroundings, or both much fainter and colder. The strong
$h_3-V$ anti-correlation argues that they are much colder. In either
case, the luminosity requirement is most easily fulfilled for our
sample as the (projected) surface brightness of disks is maximized
when viewed edge-on (assuming the central disks to be coplanar with
the large-scale galactic disks). A quantitative estimate of how
luminous (and massive) these disks must be would nevertheless be
useful, and the $K$-band images will help constrain this.

While the conventional view is that both a classical bulge and a
(separate) bar can usually be clearly identified photometrically in
face-on barred systems (at least in early-type ones), it is
questionable whether those bulges are systematically hot. Arguments
that central disks can masquerade as bulges date back at least to
\citet{k82}, and \citet{k93} reviewed much of the photometric and
kinematic evidence available then. Among others, \citet{ss96} and
\citet{ebgb03} more recently argued for the presence of significant
central disks in bulges, using high-quality observations. In fact,
much recent work shows that bulges are often not prominent and can
have rather shallow light profiles, with Sersic index $n$ ranging from
roughly $0.5$ to $2.5$ \citep[e.g.][]{apb95,j96,mch03,bgdp03}, as well
as young stars and bar or spiral structures
\citep[e.g.][]{cszm97,csm98,es02,es03}. See \citet{k01} for similar
arguments.

Those observations raise the interesting possibility that the
decoupled central stellar disks identified in most of the sample
galaxies are in fact the so-called bulges seen in face-on barred
galaxies. While their scale seems rather small (i.e.\ typically equal
to the width of the bar), they are (quasi-)axisymmetric, have similar
stellar kinematics, and also appear to have similar luminosity
profiles. Indeed, even simple optical light profiles derived by
summing our data in wavelength (not shown) show that the central light
peak is close to exponential and typically extends over the first
velocity hump only (approximately to the radius where the box/peanut
shape is maximum, although this varies somewhat as a function of
viewing angle). This again argues against merger-driven growth for
those bulges and suggests that many classical bulges may well be
concentrated disks (i.e.\ non-exponential disks), with a structure
closer to truncated cylinders rather than spheroids
\citep[see][]{k93}. Axisymmetric ``bulges'' would then grow at the
center of bars, rather than out of their destruction.

The implications from the above is that the central peak in the light
profile, normally associated with the bulge in face-on systems, is
kinematically cold and contained {\em within} the vertically extended
(thick) region of edge-on systems (i.e.\ the B/PS bulge), itself only
a portion of the bar (which extends to the turnover radius of the
rotation curve). This clearly remains speculative without good quality
photometry, but those issues will be explored further with the
$K$-band observations presented in \citet{ababdf04} and
\citet{aab04}. Similar ideas are expressed by \citet{ebgb03} in the
specific cases of \objectname{NGC~2787} and \objectname{NGC~3945}.
\subsection{Secular Evolution\label{sec:evolution}}
The N-body simulations discussed by \citeauthor{ba04}
(\citeyear{ba04}; but see also \citealt{a02,a03}) undergo substantial
bar-driven evolution as the disk transfers angular momentum to the
halo. The bar and its associated kinematic signatures grow in length
and strength with time, but there is no evidence that the central
parts circularize. It is thus doubtful whether purely stellar
processes could give rise to the central $h_3-V$ anti-correlations and
the (quasi-)axisymmetric central stellar disks argued for above.

On the other hand, our results fit nicely in a scenario where the
identified central stellar disks form through bar-driven gas
inflow. Because of the torques and shocks they generate, bars are very
efficient at driving gas toward their centers (assuming there is a
sufficient reservoir of material at large radii; see e.g.\
\citealt{a92b,fb93,fb95}). The gas tends to accumulate near the inner
Lindblad resonance (ILR, if present), often vigorously forming stars
and giving rise to a highly visible nuclear ring
\citep[e.g.][]{wfmmb95,mf97}. The conditions and behavior of the gas
at and within the ILR are still under debate (but see, e.g.,
\citealt{es00}; \citealt{sh02} and references therein), but this is
unimportant for our discussion as long as a sufficient mass is
accumulated to circularize the potential.

Once the potential is (quasi-)axisymmetric, $h_3$ and $V$ will
anti-correlate (but the $V$ and $\sigma$ signatures will
persist). Furthermore, any star formation in the gas disk will enhance
the stellar kinematic features and keep the central stellar disk cold,
negating the effects of possible heating mechanisms. This is likely as
there is overwhelming evidence for star forming central gas disks
(nuclear spirals) in most of the sample galaxies with extended
emission \citep{bf99}. In fact, a visual inspection shows that the
strength of these ionized-gas disks loosely correlates with the
strength of the central $h_3-V$ anti-correlation. For the galaxies
with no emission, we must assume either that the gas is not ionized
or, more likely, that the reservoir of material has been exhausted and
star formation in the central disks has ceased. This is consistent
with the fact that no central $\sigma$ minimum is seen in any of the
gas-poor galaxies, suggesting that their central disks are
kinematically hotter, and thus presumably older, than gas-rich systems
(\objectname{IC~4767} has strong emission but it is confined to the
inner few arcseconds).

Current or recent star formation in the central disks is further
supported by our experimentations with emission line subtraction
(\S~\ref{sec:quality}). Indeed, while the results for the two stellar
templates (K1~III and G5~IV) are identical, this is not the case when
using a single template and comparing the kinematics derived from the
entire spectra (including H$\beta$) or the Mg~$b$ region only. Then,
the $V$ and $h_3$ profiles are essentially identical, but as shown by
Figure~\ref{fig:sigma_test} the Mg~$b$ dispersions are systematically
higher for about one-third of the sample. As young stars contribute
preferentially to H$\beta$, the older stars appear kinematically
hotter. The effect extends mainly over the rapidly rising part of the
rotation curve (or at most to the end of the first plateau), where the
central $\sigma$ minima and the $h_3-V$ anti-correlations are
observed. As the first velocity hump is originally caused by the
projection of the $x_1$ orbits (perceived excess rotation; see
\citealt{ba04}), this implies that the central disks are always
contained within the width of the bar, exactly what is expected if
they are fed from it (shocks occur along the leading edges of the bar;
see e.g.\ \citealt{a92b}).

Our data thus seem perfectly consistent with a scenario where bars
give rise to both the observed B/PS bulges (through bar-buckling) and
decoupled, nearly axisymmetric central stellar disks (through gas
inflow and subsequent star formation). Adding gas and star formation
to the simulations of \citet{ba04} or a detailed comparison with
similar existing models \citep[e.g.][]{hs94,fb95} would thus be
extremely valuable. Given that at least $40-45\%$ of bulges across all
morphological types are B/PS \citep{ldp00a}, our results have
important implications for our understanding of the structure and
evolution of all bulges. Bulges here appear to be made up mostly of
disk material which has acquired a large vertical extent through
vertical instabilities. Together with the central disks, they are thus
consistent with a formation dominated by bar-driven secular evolution
processes rather than merging.

We note however that while bar-driven secular evolution scenarios are
often invoked to argue for a gradual drift of late-type bulges toward
earlier types, this aspect of the scenarios is not effectively probed
here. We have no way to tell whether the bars detected here will later
on be destroyed by the sufficient accumulation of gas in their centers
(e.g.\ \citealt{hs96}; \citealt{nsh96} and references therein), thus
contributing to the growth of a more axisymmetric spheroidal bulge, or
whether this process may have already occurred in some of the
galaxies. The kinematics presented here, although limited to the
equatorial plane, appears perfectly consistent with that expected from
disk material only. We do not {\em require} the presence of an
additional ``classical'' (i.e.\ both axisymmetric and hot) bulge in
addition to the (thick) bar detected. Whether we can rule out the
presence of a rounder and more pressure supported component with our
observations is harder to answer, and it would presumably require
detailed modeling of individual objects. Such a component may be
consistent with the velocity dispersion peaks observed in some objects
(particularly when using Mg~$b$ only), but it tends to disagree with
the flat $\sigma$ profiles (or local central $\sigma$ minimum)
observed in others. $K$-band imaging and kinematics at large galactic
heights are better suited to constrain the additional presence of a
classical bulge and will be discussed elsewhere
(\citealt{ababdf04,aab04}; Zamojski et al.\ 2004, in preparation). We
note however that \citet{ebgb03} argued for the presence of both a
concentrated inner disk and a (much) smaller classical bulge in the
two galaxies they studied (\objectname{NGC~2787} and
\objectname{NGC~3945}).

We also note that there is limited evidence from our sample for
so-called hybrid bar/bulge formation scenarios \citep{mwhob95},
whereby a bar is first excited by an interaction
\citep*[e.g.][]{n87,gca90} and then buckles. Indeed,
\objectname{NGC~128}, \objectname{NGC~6771}, and
\objectname{ESO~597-G036} all have strong peanut-shaped bulges (strong
bars) and presumably interacting close-by companions. Furthermore,
\objectname{NGC~128}, \objectname{NGC~1596}, and \objectname{NGC~3203}
have counterrotating ionized-gas disks (Bureau \& Chung 2004, in
preparation), clear evidence for the accretion of external material.
\subsection{Literature Data\label{sec:literature}}
A number of studies are now available in the literature containing
extended and reliable stellar kinematic profiles of galaxies (many
including $h_3$ and $h_4$). While they generally focus on early-type
systems, we review below studies including a large number spirals
and/or lenticulars, preferably viewed edge-on.

\citet{f97} presents high quality stellar kinematic of $18$ S0s. $9$
are exactly or approximately edge-on, $6$ of which appear boxy on the
isophotes presented (\objectname{NGC~1461}, \objectname{NGC~2560},
\objectname{NGC~4026}, \objectname{NGC~4036}, \objectname{NGC~4111},
and \objectname{NGC~4251}). All of those show the characteristic
kinematic bar signatures seen in the current sample
(\S~\ref{sec:diagnostics} and \citealt{ba04}): double-hump rotation
curve, velocity dispersion profile with a flat top and/or shoulder (or
secondary maximum), and $h_3-V$ correlation at moderate radii. Two of
the non-boxy galaxies show kinematic bar signatures
(\objectname{NGC~4350} and \objectname{NGC~4762}), but they also show
photometric evidence of a bar (a plateau in the major-axis surface
brightness profile at moderate radii; see \citealt{ldp00b} and
\citealt{ba04}). They thus most likely harbor a bar seen end-on (but
see \citealt{w94} for \objectname{NGC~4762}). The third non-boxy
galaxy (\objectname{NGC~5666}) shows neither kinematic nor photometric
bar signatures. All the presumably barred galaxies also show an
$h_3-V$ anti-correlation in the center, as observed
here. \citeauthor{f97}'s (\citeyear{f97}) results are thus entirely
consistent and supportive of ours. Based on photometric evidence, he
in fact suggested that bars could be at the origin of the $h_3$ slope
reversals observed, but he lacked the diagnostics of \citet{ba04} to
confirm it. $3$ of the above galaxies also show a central ionized-gas
disk and $3$ a local minimum in the minor-axis $\sigma$ profile,
further supporting the presence of a thin, rapidly rotating stellar
(and sometimes gas) disk at the center of the objects. It would be
interesting to see if the ionized-gas kinematics shows the same kind
of bar signatures predicted by \citet{ab99} and observed by
\citet{mk99} and \citet{bf99}.

Although we prefer to delay the discussion of photometric papers to
\citet{ababdf04}, \citet{ss96} discuss the photometry of $16$ nearly
edge-on S0s along with the stellar kinematics of $10$. They derive the
major-axis disk profile of each object by subtracting a pure
ellipsoidal model constrained above the equatorial plane. They almost
systematically identify an outer disk with an inner cut-off, but also
a number of steep inner disks. Although no explanation is offered for
either structure, those results are expected in the context of
bar-driven secular evolution scenarios. $12$ galaxies have a light
profile plateau at intermediate radii, typical of a bar viewed
edge-on. All the double-disks are observed in galaxies with such a
plateau (i.e.\ a bar), as are all the rings identified. The latter
also systematically occur at the end of the plateau and the turnover
in the rotation curve, as expected for inner rings and bars
\citep[e.g.][]{ba04}. Among the $8$ objects with a light plateau and
stellar kinematics (\objectname{NGC~2549}, \objectname{NGC~2732},
\objectname{NGC~3098}, \objectname{NGC~4026}, \objectname{NGC~4111},
\objectname{NGC~5308}, \objectname{NGC~5422}, and
\objectname{NGC~7332}), $6$ have a double-hump velocity profile
characteristic of bars (all also have a double-disk structure) and
only $2$ do not (none have a double-disk structure). Given the poor
spatial sampling of the data and the absence of error bars or seeing
measurements, this is perfectly consistent with a one-to-one
relationship between bars and light plateaus. Therefore, while
\citeauthor{ss96}'s (\citeyear{ss96}) inner disks can be assimilated
with the central stellar disks argued for here (i.e.\ the central
region where $h_3$ and $V$ anti-correlate), their arguments for
truncated outer disks are questionable. As one moves from the outside
in, we see no evidence of a disk truncation, but rather simply the
beginning of the bar and an increase in the scaleheight of the disk
(the B/PS bulge and $h_3-V$ correlation region; see
\citealt{aab04}). No separate (classical) bulge component is required
either.

\citet{egcflpw01} recently obtained high spatial resolution stellar
kinematics of $4$ intermediate-type barred spiral galaxies, each with
a Seyfert nucleus, secondary bar, and moderate inclination
(\objectname{NGC~1097}, \objectname{NGC~1365}, \objectname{NGC~1808},
and \objectname{NGC~5728}). All $4$ galaxies show a double-hump
rotation curve and $3$ a central $\sigma$ minimum. Simple dynamical
models require the presence of decoupled central (and weakly
non-axisymmetric) stellar disks, presumably fueled by bar-driven gas
inflow and star formation. This scenario is identical to the one
proposed here. Interestingly, the central disks required have similar
masses and radii to the small (classical) spheroidal bulges assumed.

It would also be interesting to know if the so-called ``double-wave''
rotation curves, reported by \citet{b89} along the bar of many SB0s
($7$ of the $10$ objects studied), could be related to the double-hump
velocity profiles observed here (or to the presence of decoupled
centers). A generic interpretation is harder in this case as the
orientation of each object differs (orientation of the bar with
respect to the line-of-nodes and inclination), but an explanation
relying on the axial ratio variation of the $x_1$ orbits again appears
likely (comparing to the 2D model of \citealt{ss87}; see also
\citealt{ba99,ba04}). A comparison with current 3D N-body simulations
would be straightforward.

Going beyond kinematics, stellar populations also offer an important
tool to constrain galaxy formation scenarios. The absence of
correlations between velocity dispersion and linestrength gradients
(at least in some bulge samples; e.g.\ \citealt*{ffi96,f97}) argues
against pure dissipational collapse \citep{l76,c84} and other
arguments limit the influence of accretion and/or
merging. Unfortunately, detailed predictions for the stellar
populations are still largely inexistent for bar-driven secular
evolution scenarios. Naively, one would expect bars to smooth out
population gradients due to substantial radial (and vertical) motions,
but any central star formation and subsequent stellar migration
(mixing) will greatly complicate this picture and may actually enhance
gradients in the center (for early models, see
\citealt*{fbk94,fb95}). This seems to be borne out by observations
(see, e.g., \citealt{mr94} and \citealt*{zkh94} for the gas;
\citealt{ffi96} for the stars), and NGC~4594 (Sombrero galaxy;
\citealt{ebmp96}) and NGC~7332 \citep{fetal04} offer excellent
examples where largely successful attempts have been made to interpret
both the gaseous/stellar kinematics and the stellar populations
coherently within a bar-driven secular evolution
scenario. Nevertheless, detailed chemo-dynamical models of barred
galaxies taking into account all of the above effects are urgently
needed.

Integral-field spectrographs with a large field-of-view and sufficient
wavelength coverage offer the best potential to constrain the
intermediate-scale structure of galaxies and (in)validate formation
scenarios. In this respect, the {\tt SAURON} survey is particularly
relevant, having mapped the ionized-gas distribution and kinematics,
the stellar kinematics, and the linestrength indices (age and
metallicity) of over $70$ ellipticals, lenticulars, and Sa bulges to
one effective radius \citep[see][]{zetal02,eetal04}. The 2D
information confirms that a number of galaxies possess a structure
similar to that argued for here, i.e.\ a cold and thin central
stellar disk (first $V$ increment, central $\sigma$ minimum, and
$h_3-V$ anti-correlation) embedded in a much larger and thicker
triaxial component (first $V$ plateau and $h_3-V$
correlation). Central ionized-gas disks are also often cospatial with
the central stellar disks. The best examples are NGC~4526 and NGC~3623
\citep{zetal02,betal02a,betal02b}, but there are numerous others
\citep{fetal04,eetal04}.
%
%
\section{Conclusions\label{sec:conclusion}}
The (projected) morphology of B/PS bulges departs strongly from that
of classical, spheroidal bulges \citep[e.g.][]{s93b}, implying an even
more distorted 3D structure, but their high incidence ($\gtrsim45\%$
of all bulges; see \citealt{ldp00a}) makes them more than simple
curiosities and a widespread mechanism must be able to account for
their structure and dynamics. By studying the stellar kinematics of
the $30$ edge-on spiral galaxies of \citet{bf99}, we have attempted
here to discriminate between the two leading formations scenarios:
accretion \citep[e.g.][]{bp85,r88} and bar-buckling
\citep[e.g.][]{cs81,cdfp90}. The latter is particularly attractive in
view of currently popular secular evolution scenarios, which argue
that bars may drive bulges along the Hubble sequence (e.g.\
\citealt{fb95}; \citealt{nsh96} and references therein).

Our results are summarized in Table~\ref{tab:class}. $22$ of $24$
galaxies with a B/PS bulge show clear kinematic bar signatures
($\approx90\%$), while only $2$ of the $6$ control sample galaxies do
and are probably transition objects. We thus confirm the results of
\citet{bf99} for galaxies with extended ionized-gas and, for the first
time, extend those results to gas-poor B/PS bulges (S0s). The
kinematic evidence in favor of a close relationship between thickened
bars and B/PS bulges is thus now overwhelming across all morphological
types (S0--Sbc), although we still have not probed the potentials out
of the equatorial plane. Zamojski et al.\ (2004, in preparation) shall
discuss the stellar kinematics of a few objects at large galactic
heights.

Most objects show kinematic profiles in agreement with those predicted
from N-body simulations of barred disks \citep{ba04}, including the
expected relationship between bulge morphology and bar viewing
angle. In particular, most galaxies with a B/PS bulge have i) a
double-hump rotation curve with a dip or plateau at moderate radii,
ii) a rather flat central velocity dispersion profile with a secondary
peak or plateau and (in about $40\%$ of cases) a local central
minimum, and iii) an $h_3$ profile correlated with $V$ over the
expected bar length, with up to three slope reversals. As expected if
caused by a bar, the strengths and extents of those features are
correlated. Our $h_4$ profiles are generally too noisy to be of much
use.

A number of differences with the simulations also exist, most of which
appear related to the past or current presence of gas. In particular,
the kinematic profiles of most galaxies show an anti-correlation of
$V$ and $h_3$ in the very center, over a region corresponding to the
rapidly rising part of the rotation curve (or at most the first
plateau) and encompassing the central light peak and central $\sigma$
minimum (if present). Being cospatial with observed ionized-gas disks
in gas-rich galaxies \citep{bf99}, this kinematic behavior is most
likely due to cold quasi-axisymmetric central stellar disks, which we
argue have decoupled due to bar-driven gas inflow, and may have been
enhanced by subsequent star formation. B/PS bulges are thus entirely
consistent with the predictions of most bar-driven secular evolution
scenarios \citep[e.g.][]{hs94,fb95}, whereby bars can simultaneously
account for the creation of a large vertically extended component out
of disk material and for the accumulation of mass in their centers
(possibly associated with the so-called bulges of face-on systems),
bypassing the need for merging or accretion of external material.

Lastly, we reiterate that $h_3$ is a very useful tracer of the
structure and dynamics of galaxies. As suggested by \citet{ba04}, the
correlation or anti-correlation of $h_3$ and $V$ appears to be a
reliable diagnostics of the intrinsic shape, at least for highly
inclined disks.
%
%

\acknowledgements Support for this work was provided by NSF grant
AST~00-98249 to Columbia University and by NASA through Hubble
Fellowship grant HST-HF-01136.01 awarded by the Space Telescope
Science Institute, which is operated by the Association of
Universities for Research in Astronomy, Inc., for NASA, under contract
NAS~5-26555. M.B.\ wishes to thank K.\ C.\ Freeman for his involvement
in the early stages of this project and E.\ Athanassoula, A.\ Bosma,
and E.\ Emsellem for useful discussions. The Digitized Sky Surveys
were produced at the Space Telescope Science Institute under
U.S. Government grant NAG W-2166. The images of these surveys are
based on photographic data obtained using the Oschin Schmidt Telescope
on Palomar Mountain and the UK Schmidt Telescope. The plates were
processed into the present compressed digital form with the permission
of these institutions. This research has made use of NASA's
Astrophysics Data System (ADS) Bibliographic Services and the
Lyon-Meudon Extragalactic Database (LEDA; http://leda.univ-lyon1.fr).
%
%

%
\clearpage
%
%
%
\figcaption[chung.fig1.eps]{Stellar kinematics of galaxies with a
B/PS bulge and no or confined emission. From top to bottom, the panels
show an optical image of the galaxy from the Digitized Sky Surveys
(with isophotal contours), and the registered $V$, $\sigma$, $h_3$,
and $h_4$ profiles along the major-axis. The spatial scale varies from
galaxy to galaxy but the kinematic limits are fixed. The measurements
were folded about the center and the errors represent half the
difference between the approaching and receding sides. Measurements
are mirrored and denoted by a cross without an error bar when
available from one side only. The data were binned to $S/N\ge5$ per
spatial and spectral element for $V$ and $\sigma$ and $S/N\ge10$ for
$h_3$ and $h_4$. The final $S/N$ are thus at least $7$ and $14$,
respectively.\label{fig:kin_noem}}
%
%
\figcaption[chung.fig2a.eps]{Same as Figure~\ref{fig:kin_noem} but
for the galaxies with a B/PS bulge and extended emission (main sample
of \citealt{bf99}).\label{fig:kin_em}}
%
%
\figcaption[chung.fig3.eps]{Same as Figure~\ref{fig:kin_noem} but for
the galaxies of the control sample (control sample of
\citealt{bf99}).\label{fig:kin_control}}
%
%
\figcaption[chung.fig4.eps]{Velocity dispersion of all sample galaxies
derived using the entire spectral range (solid circles) and the Mg~$b$
region only (crosses). Errors bars are not plotted for
clarity.\label{fig:sigma_test}}
%
%
\figcaption[chung.fig5.eps]{Comparison of our stellar measurements with
published stellar kinematics. Data from this paper are shown as filled
circles and literature data as asterisks and boxes. Error bars are
drawn when available and references are available in the
text.\label{fig:comp_stellar}}
%
%
\figcaption[chung.fig6.eps]{Same as Figure~\ref{fig:comp_stellar} but
for the published gas kinematics.\label{fig:comp_gas}}
%
%
\figcaption[chung.fig7.eps]{Rotational and pressure support for all
sample galaxies. The dimensionless parameter $V/\sigma$ is shown as a
function of projected radius. Crosses without an error bar denote data
available from one side only.\label{fig:v/sigma}}
%
%
\figcaption[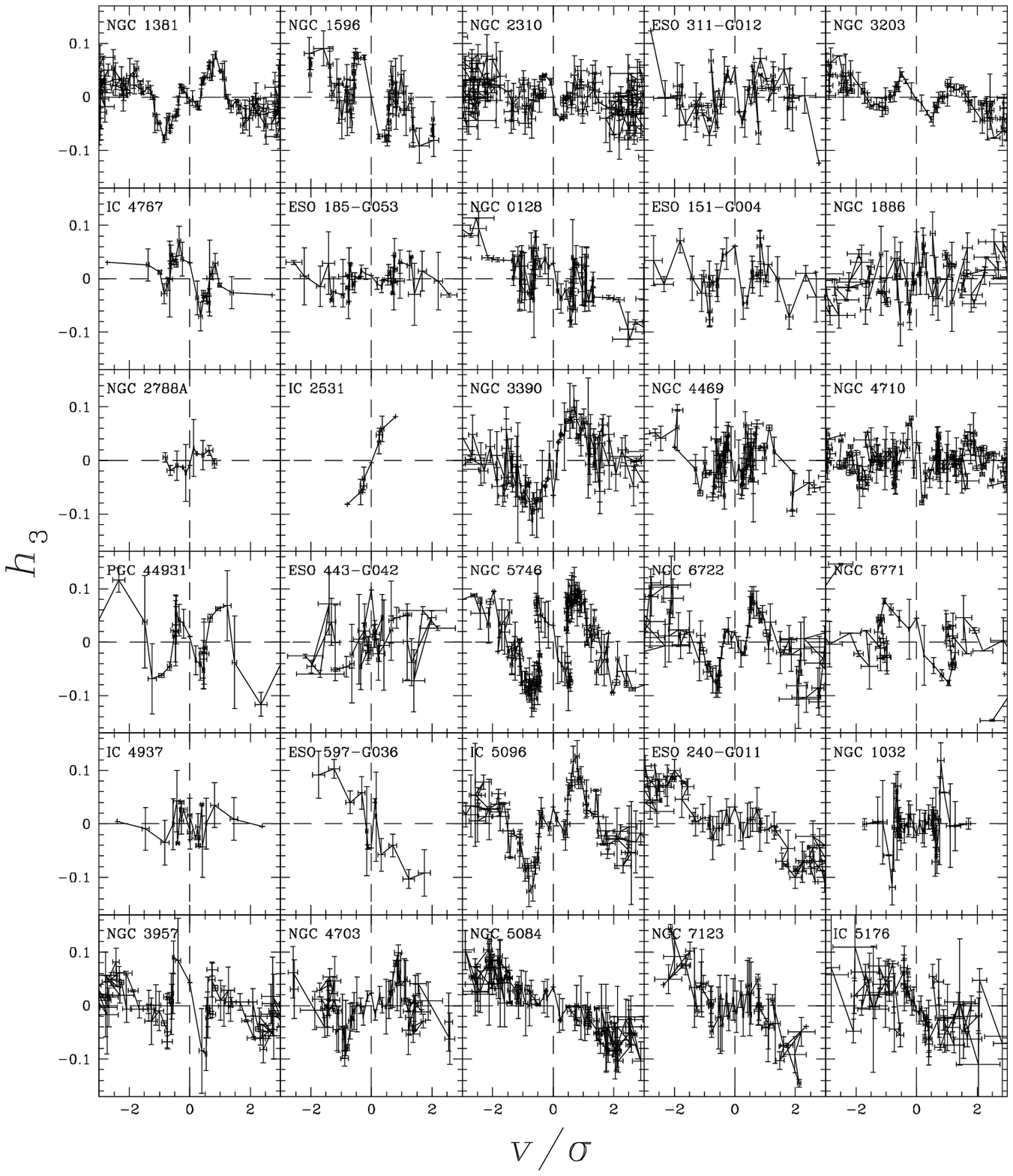]{Velocity profile asymmetries for all
sample galaxies. The LOSVD asymmetry (skewness) $h_3$ is shown as a
function of $V/\sigma$. Data with $S/N\ge10$ were used here for both
$V$, $\sigma$, and $h_3$. Crosses without an error bar denote data
available from one side only. The points are connected as a function
of projected radius.\label{fig:h3}}
%
%
\figcaption[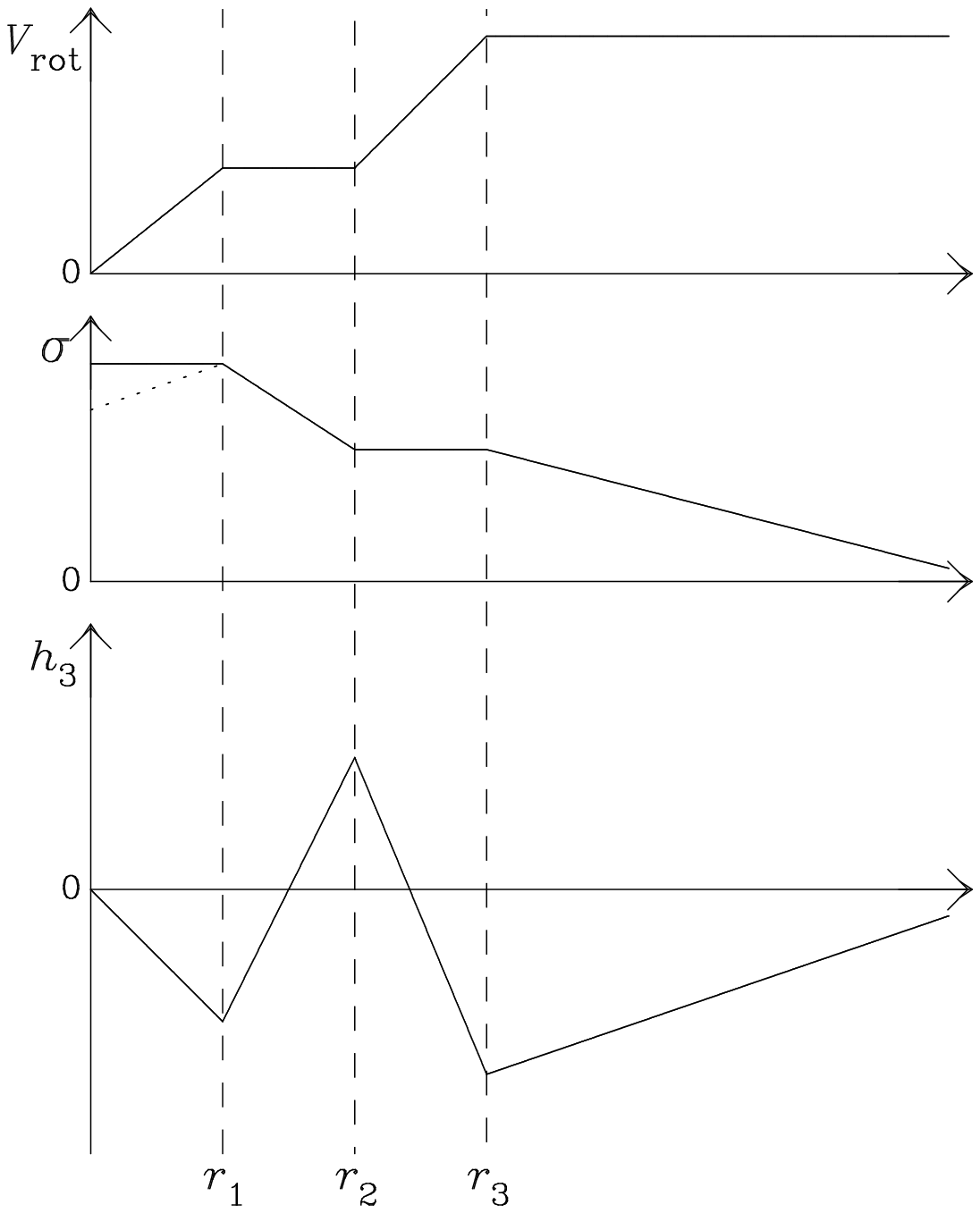]{A schematic view of the kinematic
profiles for $V$, $\sigma$, and $h_3$ derived for most sample
galaxies.\label{fig:schem_diagram}}
%
%
%
\begin{deluxetable}{lrrlrrrrlc}
\tabletypesize{\tiny}
\tablewidth{0pt}
\tablecaption{General properties of sample galaxies.\label{tab:prop}}
\tablehead{Galaxy & R.A.(2000) & Dec.(2000) & Type & $B_{\rm T}$ &
$D_{25}$ & $V_{\rm hel}$~~~ & $M_B^{\rm c}$ & Environment & Note \\
 & $^{~h~~~m~~~s}$ & $^{~~~\circ~~~\arcmin~~~\arcsec}$ & & ~mag &
arcmin & km~s$^{-1}$ & ~~~~mag & \\
{~~~(1)} & 
\multicolumn{1}{c}{(2)\tablenotemark{a}} &
\multicolumn{1}{c}{(3)\tablenotemark{a}} &
{~~(4)\tablenotemark{b}} & 
\multicolumn{1}{c}{(5)\tablenotemark{a}} & 
\multicolumn{1}{c}{~~(6)\tablenotemark{a}} & 
\multicolumn{1}{c}{(7)\tablenotemark{a}} & 
\multicolumn{1}{c}{(8)\tablenotemark{a}} & 
\multicolumn{1}{c}{(9)\tablenotemark{c}} & 
\multicolumn{1}{c}{(10)\tablenotemark{c}} \\}
\startdata
\noalign{\vspace{0.15cm}}
\multicolumn{10}{l}{\bf Galaxies with no or confined emission}\\
NGC 1381     & 03 36 31.7 & -35 17 43 & SA0          & 12.7 & 2.63 & 1776~~~ &~ -19.2 & Cluster     & \nodata  \\
NGC 1596     & 04 27 37.8 & -55 01 37 & SA0          & 12.0 & 3.89 & 1509~~~ &~ -19.3 & Group       & \nodata  \\
NGC 2310     & 06 53 53.8 & -40 51 46 & S0           & 12.6 & 4.16 & 1187~~~ &~ -18.6 & Isolated    & \nodata  \\
ESO 311-G012 & 07 47 34.0 & -41 27 07 & S0/a?        & 12.4 & 3.71 & 1130~~~ &~ -20.0 & Isolated    & \nodata  \\
NGC 3203     & 10 19 34.4 & -26 41 53 & SA(r)0$^{+}$?& 13.0 & 2.81 & 2410~~~ &~ -19.9 & Group       & \nodata  \\
IC  4767     & 18 47 41.6 & -63 24 20 & S pec        & 14.3 & 1.51 & 3544~~~ &~ -19.5 & Cluster     & \nodata  \\
ESO 185-G053 & 20 03 00.4 & -55 56 53 & SB pec       & 14.3 & 1.23 & 4475~~~ &~ -20.0 & Cluster     & \nodata  \\
\noalign{\vspace{0.15cm}}
\tableline
\noalign{\vspace{0.15cm}}
\multicolumn{10}{l}{\bf Galaxies with extended emission}\\
NGC 128      & 00 29 15.1 & +02 51 50 & S0 pec       & 12.7 & 2.81 & 4228~~~ &~ -21.4 & Group       & \nodata  \\ 
ESO 151-G004 & 00 56 07.3 & -53 11 28 & S0$^{0}$     & 14.7 & 1.31 & 7456\tablenotemark{d}~~~ &~ -20.4\tablenotemark{d} & Group       & \nodata  \\
NGC 1886     & 05 21 48.2 & -23 48 36 & Sab          & 13.8 & 3.23 & 1737~~~ &~ -19.2 & Isolated    & \nodata  \\
NGC 2788A    & 09 02 40.2 & -68 13 38 & Sb           & 13.6 & 2.88 & 4056~~~ &~ -21.5 & Cluster     & Dusty  \\
IC  2531     & 09 59 55.4 & -29 37 02 & Sb           & 12.9 & 6.76 & 2474~~~ &~ -21.6 & Cluster     & Dusty  \\
NGC 3390     & 10 48 04.4 & -31 32 02 & Sb           & 12.8 & 3.46 & 3039~~~ &~ -21.5 & Companions? & \nodata  \\
NGC 4469     & 12 29 28.0 & +08 44 59 & SB(s)0/a?    & 12.4 & 3.46 &  576~~~ &~ -17.8 & Cluster     & \nodata  \\
NGC 4710     & 12 49 38.9 & +15 09 57 & SA(r)0$^{+}$ & 11.9 & 4.89 & 1324~~~ &~ -19.8 & Cluster     & \nodata  \\
PGC 44931    & 13 01 49.5 & -08 20 10 & Sbc          & 14.2 & 2.81 & 3804~~~ &~ -21.1 & Isolated    & \nodata  \\
ESO 443-G042 & 13 03 29.9 & -29 49 36 & Sb           & 13.9 & 2.88 & 2912~~~ &~ -20.6 & Companions? & Dusty  \\ 
NGC 5746     & 14 44 55.9 & +01 57 17 & SAB(rs)b?    & 11.4 & 6.91 & 1720~~~ &~ -21.8 & Group       & \nodata  \\
NGC 6722     & 19 03 39.6 & -64 53 41 & Sb           & 13.5 & 2.88 & 5749~~~ &~ -22.2 & Isolated    & \nodata  \\
NGC 6771     & 19 18 39.6 & -60 32 46 & SA(r)0$^{+}$?& 13.6 & 2.34 & 4221~~~ &~ -20.5 & Group       & \nodata  \\
IC  4937     & 20 05 17.9 & -56 15 20 & Sb           & 14.8 & 1.86 & 2337~~~ &~ -18.6 & Cluster     & Dusty  \\
ESO 597-G036 & 20 48 15.0 & -19 50 58 & S0$^{0}$ pec & 15.2 & 0.87 & 8694~~~ &~ -20.7 & Group       & Dusty  \\
IC  5096     & 21 18 21.8 & -63 45 42 & Sb           & 13.6 & 3.16 & 3142~~~ &~ -20.7 & Companions? & \nodata  \\
ESO 240-G011 & 23 37 50.5 & -47 43 37 & Sb           & 13.4 & 4.89 & 2842~~~ &~ -21.0 & Group       & \nodata  \\
\noalign{\vspace{0.15cm}}
\tableline
\noalign{\vspace{0.15cm}}
\multicolumn{10}{l}{\bf Control sample}\\
NGC 1032     & 02 39 23.6 & +01 05 38 & S0/a         & 12.7 & 3.46 & 2722~~~ &~ -20.7 & Companions? & Dusty  \\
NGC 3957     & 11 54 01.5 & -19 34 09 & SA0$^{+}$    & 13.0 & 3.09 & 1686~~~ &~ -19.0 & Cluster     & \nodata  \\
NGC 4703     & 12 49 18.9 & -09 06 30 & Sb           & 14.0 & 2.45 & 4458~~~ &~ -21.1 & Isolated    & Dusty  \\
NGC 5084     & 13 20 16.8 & -21 49 38 & S0           & 11.5 &10.71 & 1725~~~ &~ -20.9 & Cluster     & \nodata  \\
NGC 7123     & 21 50 46.4 & -70 19 59 & Sa           & 13.6 & 2.51 & 3737~~~ &~ -20.3 & Isolated    & Dusty  \\
IC  5176     & 22 14 55.3 & -66 50 56 & SAB(s)bc?    & 13.4 & 4.36 & 1746~~~ &~ -19.6 & Companions? & \nodata  \\
\enddata
\tablenotetext{a}{Parameters from LEDA (Lyon-Meudon Extragalactic Database).}
\tablenotetext{b}{Morphological classifications from \citet{j86},
 \citet{sa87}, \citet{s87}, and \citet{kkp93}.}
\tablenotetext{c}{Notes by \citet{b98} and \citet{bf99}.}
\tablenotetext{d}{Redshift and corrected absolute B magnitude from this study.}
\end{deluxetable}
\clearpage
%
%
\begin{deluxetable}{lcccc}
\tabletypesize{\tiny}
\tablewidth{0pt}
\tablecaption{Bulge classifications and bar signatures.\label{tab:class}}
\tablehead{ & \citet{ldp00a} & \multicolumn{2}{c}{\citet{bf99}} & This work\\
& {------------------------------} &
\multicolumn{2}{c}{----------------------------------} &
{------------------------} \\ 
Galaxy & Classification code\tablenotemark{a} & (DSS) 
& Gas PVD & Stellar Kinematics \\}
\startdata
\noalign{\vspace{0.15cm}}
\multicolumn{5}{l}{\bf Galaxies with no or confined emission}\\
NGC 1381     & 2.0 & Boxy       & \nodata      & Bar \\
NGC 1596     & 4.0 & Boxy       & \nodata      & Bar \\
NGC 2310     & 2.0 & Boxy       & \nodata      & Bar \\
ESO 311-G012 & 2.0 & Boxy       & \nodata      & Bar \\
NGC 3203     & 3.0 & Boxy       & \nodata      & Bar \\
IC  4767     & 1.0 & Peanut     & \nodata      & Bar \\
ESO 185-G053 & 2.0 & Peanut     & \nodata      & Bar \\
\noalign{\vspace{0.15cm}}
\tableline
\noalign{\vspace{0.15cm}}
\multicolumn{5}{l}{\bf Galaxies with extended emission}\\
NGC 128      & 1.0 & Peanut     & Bar          & Bar  \\ 
ESO 151-G004 & 1.0 & Peanut     & Bar          & Bar  \\
NGC 1886     & 1.0 & Peanut     & Bar          & Bar  \\ 
NGC 2788A    & 1.0 & Peanut     & Bar          & Bar  \\
IC  2531     & 1.0 & Peanut     & Bar          & Bar  \\
NGC 3390     & 2.0 & Boxy       & Accretion    & Bar? \\
NGC 4469     & 1.0 & Peanut     & Axisymmetric & Bar  \\
NGC 4710     & 1.5 & Boxy       & Bar          & Bar  \\
PGC 44931    & 1.0 & Peanut     & Bar          & Bar  \\
ESO 443-G042 & 1.0 & Peanut     & Bar          & Bar? \\
NGC 5746     & 1.0 & Peanut     & Bar          & Bar  \\
NGC 6722     & 1.0 & Peanut     & Bar          & Bar  \\
NGC 6771     & 1.0 & Peanut     & Bar          & Bar  \\
IC  4937     & 1.0 & Peanut     & Bar          & Bar  \\
ESO 597-G036 & 1.0 & Peanut     & Accretion    & Bar  \\
IC  5096     & 4.0 & Boxy       & Bar          & Bar  \\
ESO 240-G011 & 4.0 & Boxy       & Bar          & Bar  \\
\noalign{\vspace{0.15cm}}
\tableline
\noalign{\vspace{0.15cm}}
\multicolumn{5}{l}{\bf Control sample}\\
NGC 1032     & 4.0 & Spheroidal & Axisymmetric & No bar\\
NGC 3957     & 3.0 & Spheroidal & Axisymmetric & Bar   \\
NGC 4703     & 4.0 & Spheroidal & Axisymmetric & Bar   \\
NGC 5084     & 4.0 & Spheroidal & Accretion    & No bar\\
NGC 7123     & 4.0 & Spheroidal & Accretion    & Bar?  \\
IC  5176     & 4.0 & Spheroidal & Axisymmetric & No bar\\
\enddata
\tablenotetext{a}{1: peanut-shaped bulge; 1.5: bulge boxy-shaped on
one side and peanut-shaped on the other; 2: box-shaped bulge; 3: bulge
is close to box-shaped, not elliptical; 4: elliptical bulge.}
\end{deluxetable}
\clearpage
%
%
\begin{figure}
\epsscale{0.82}
\plotone{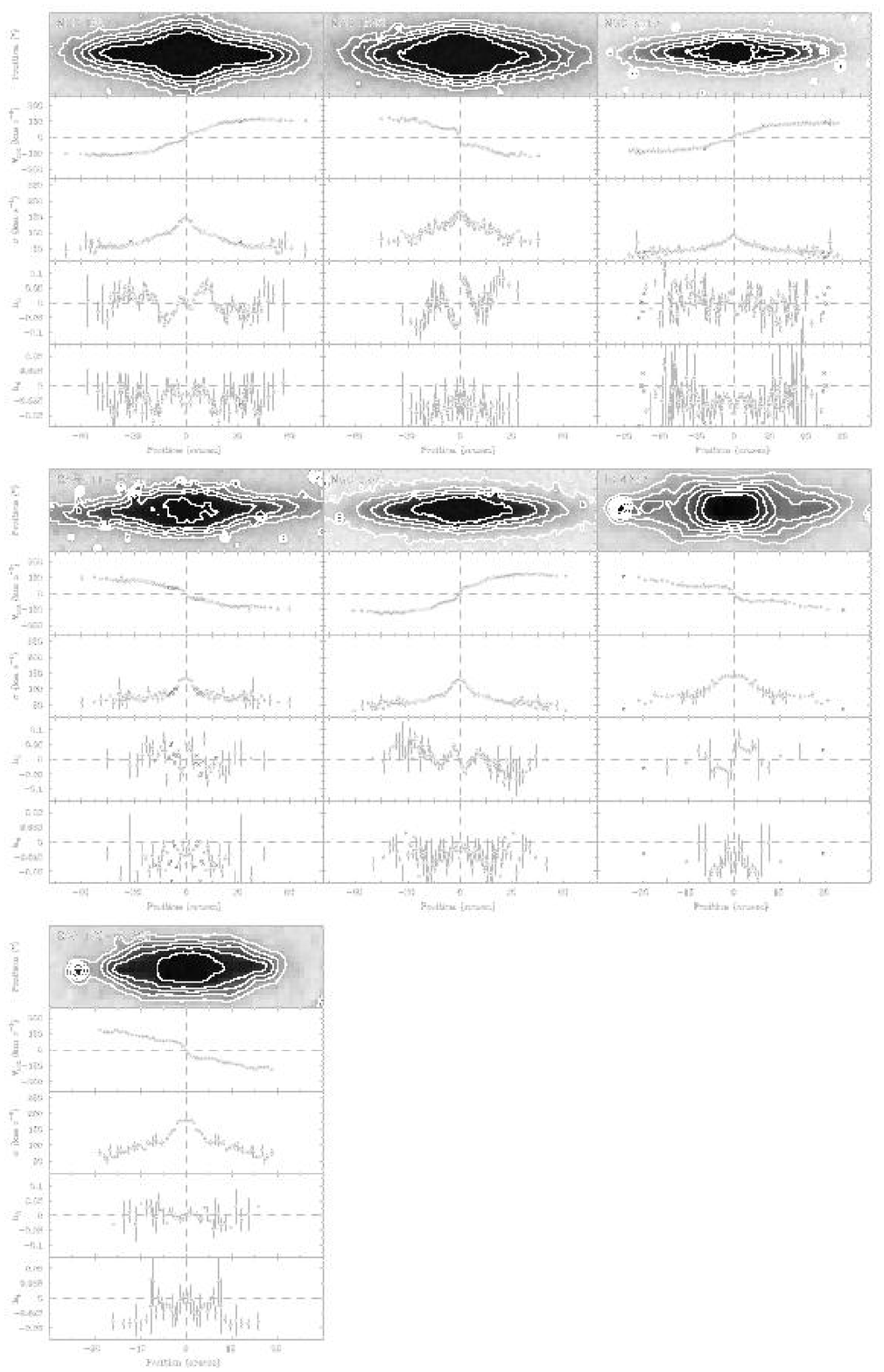}
\epsscale{1.0}
\end{figure}
\clearpage
\begin{figure}
\epsscale{0.82}
\plotone{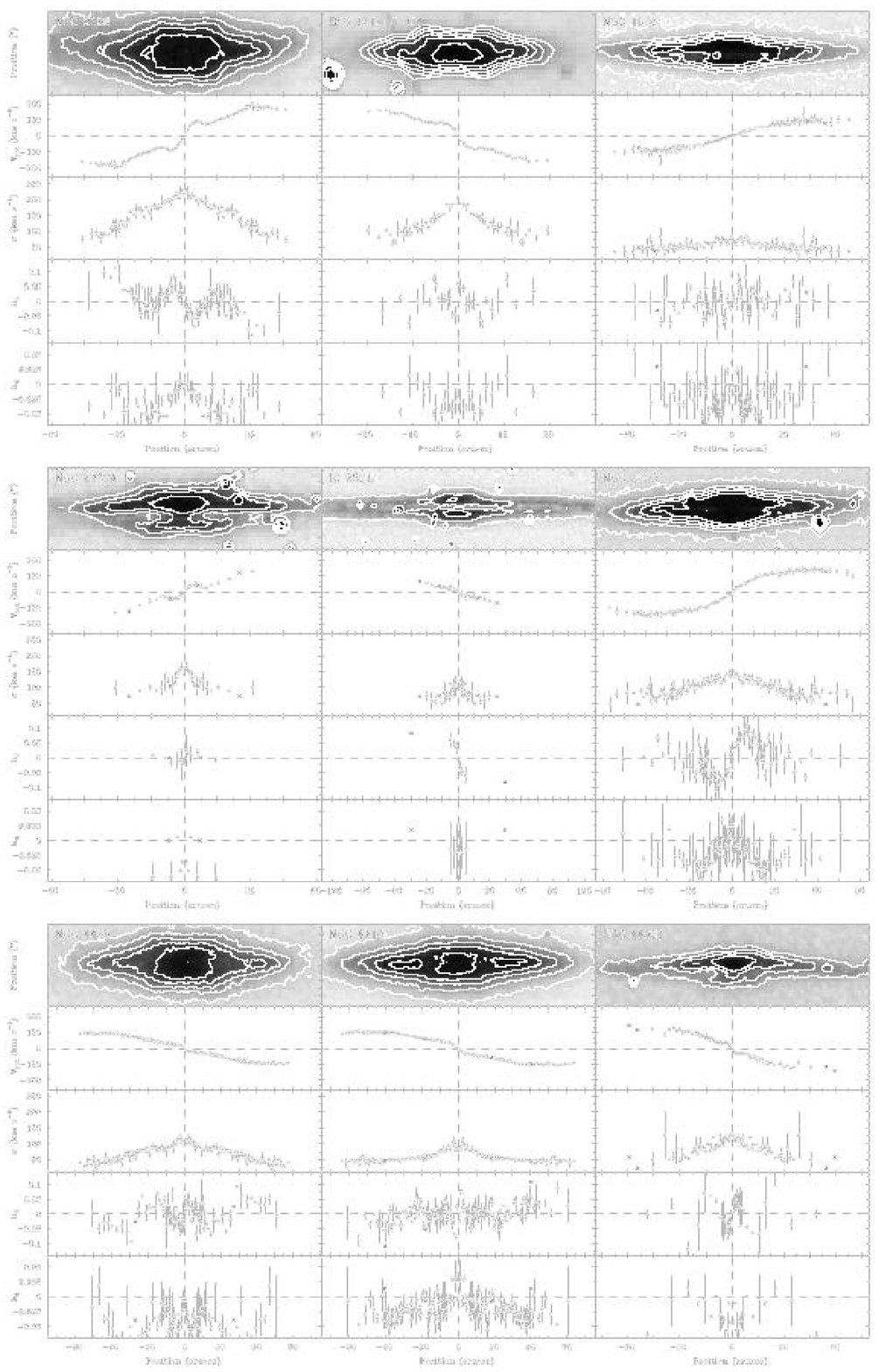}
\epsscale{1.0}
\end{figure}
\clearpage
\begin{figure}
\epsscale{0.82}
\plotone{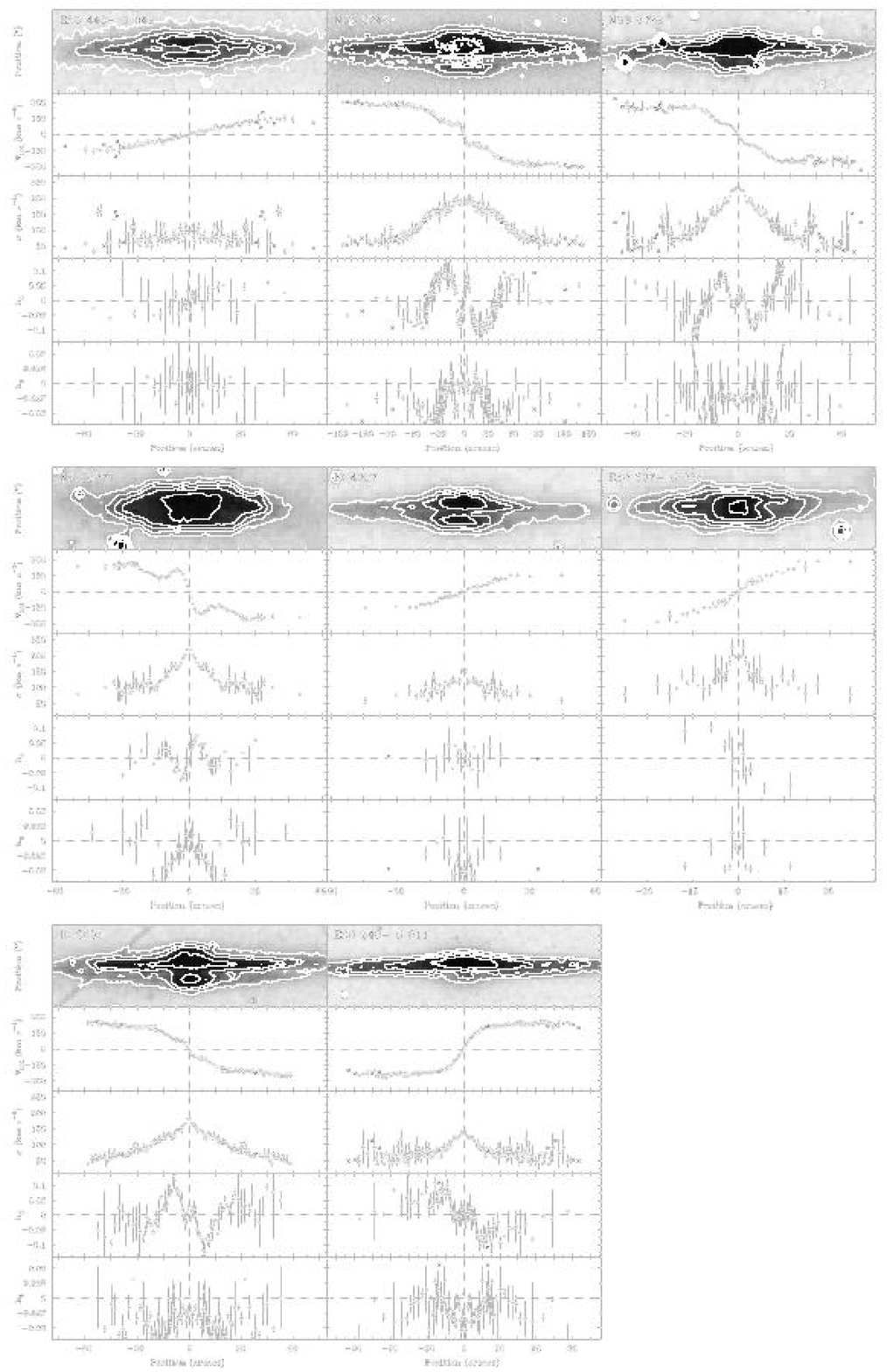}
\epsscale{1.0}
\end{figure}
\clearpage
\begin{figure}
\epsscale{0.82}
\plotone{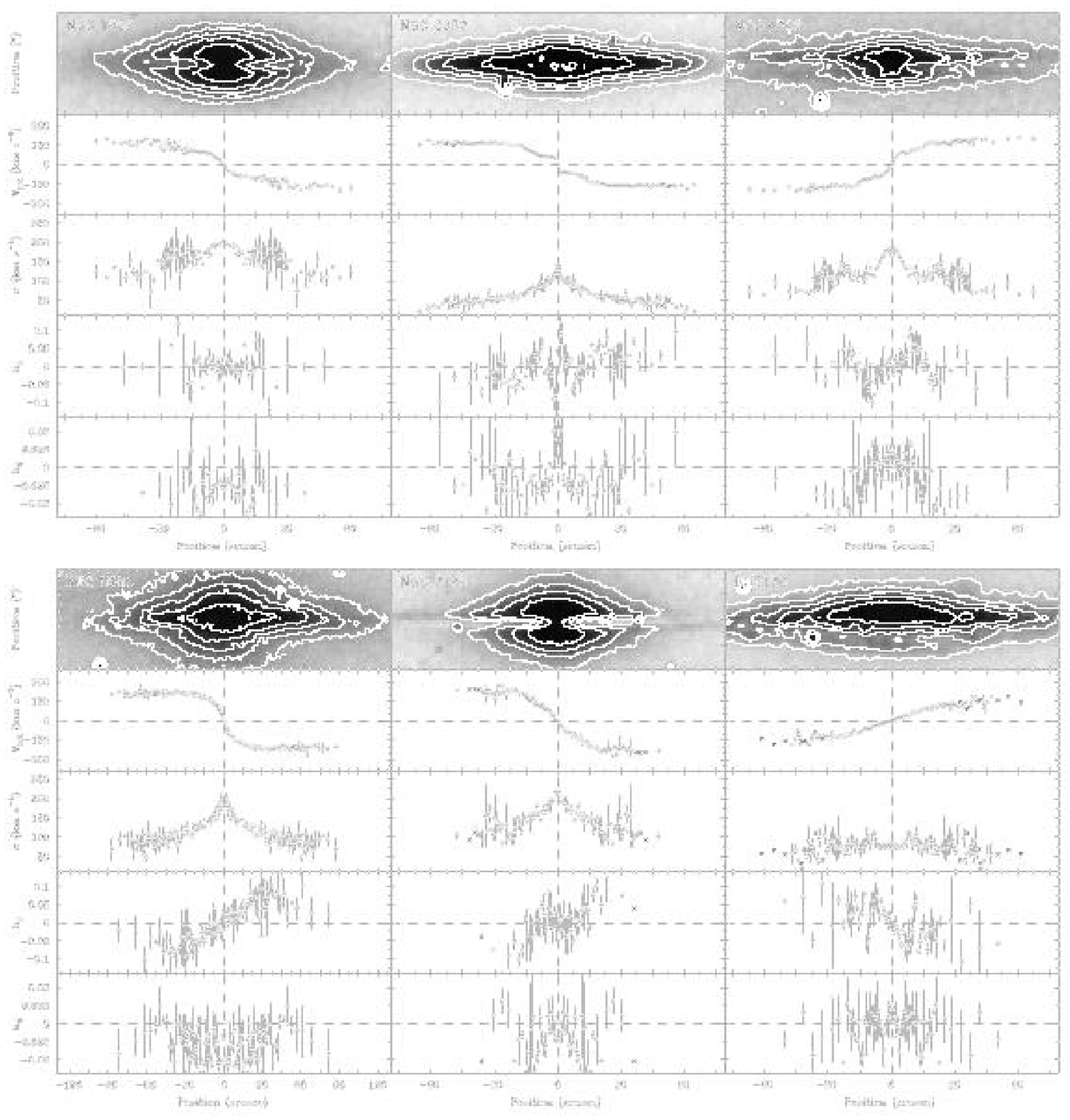}
\epsscale{1.0}
\end{figure}
\clearpage
\begin{figure}
\epsscale{0.80}
\plotone{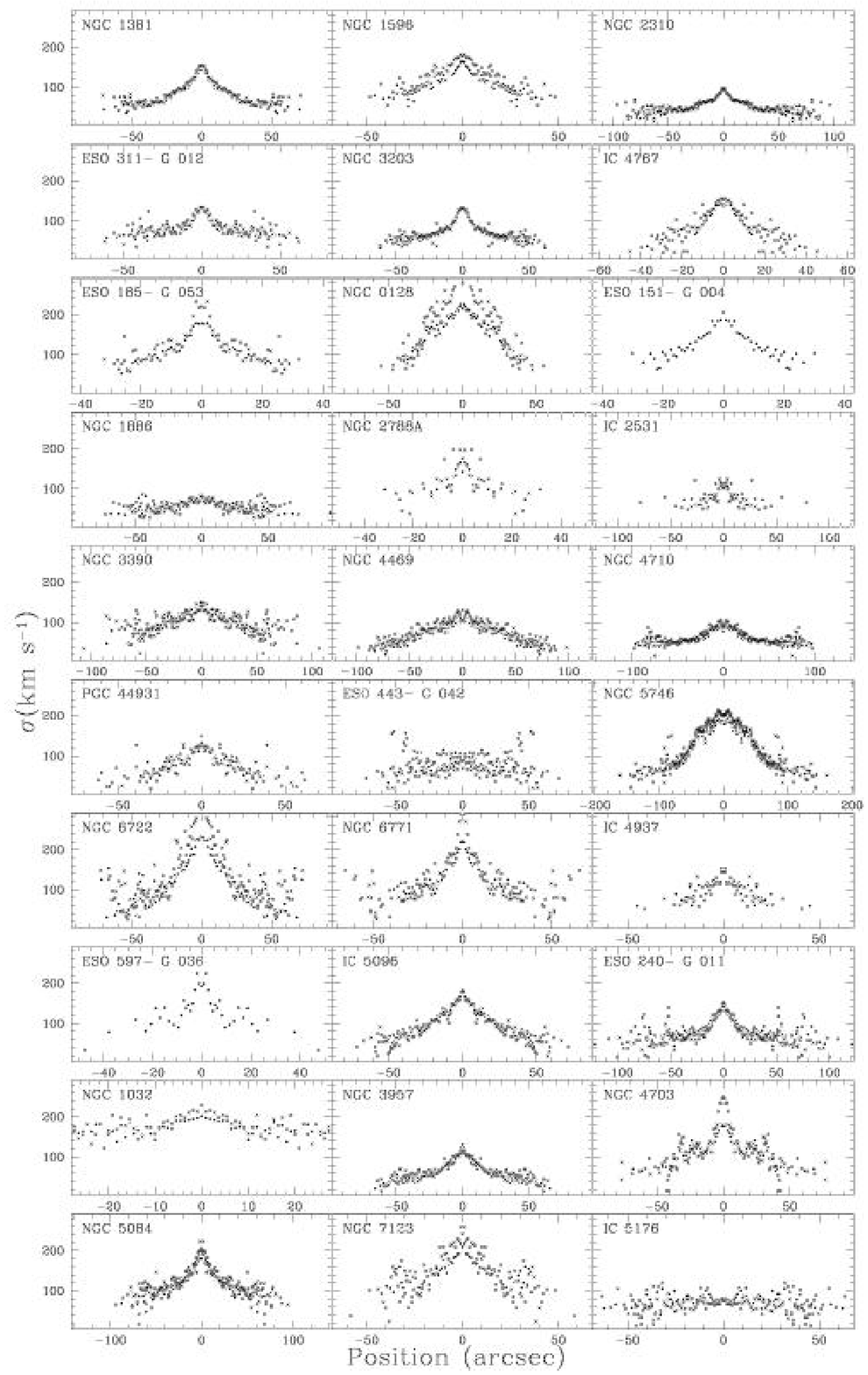}
\epsscale{1.0}
\end{figure}
\clearpage
\begin{figure}
\epsscale{1.0}
\plotone{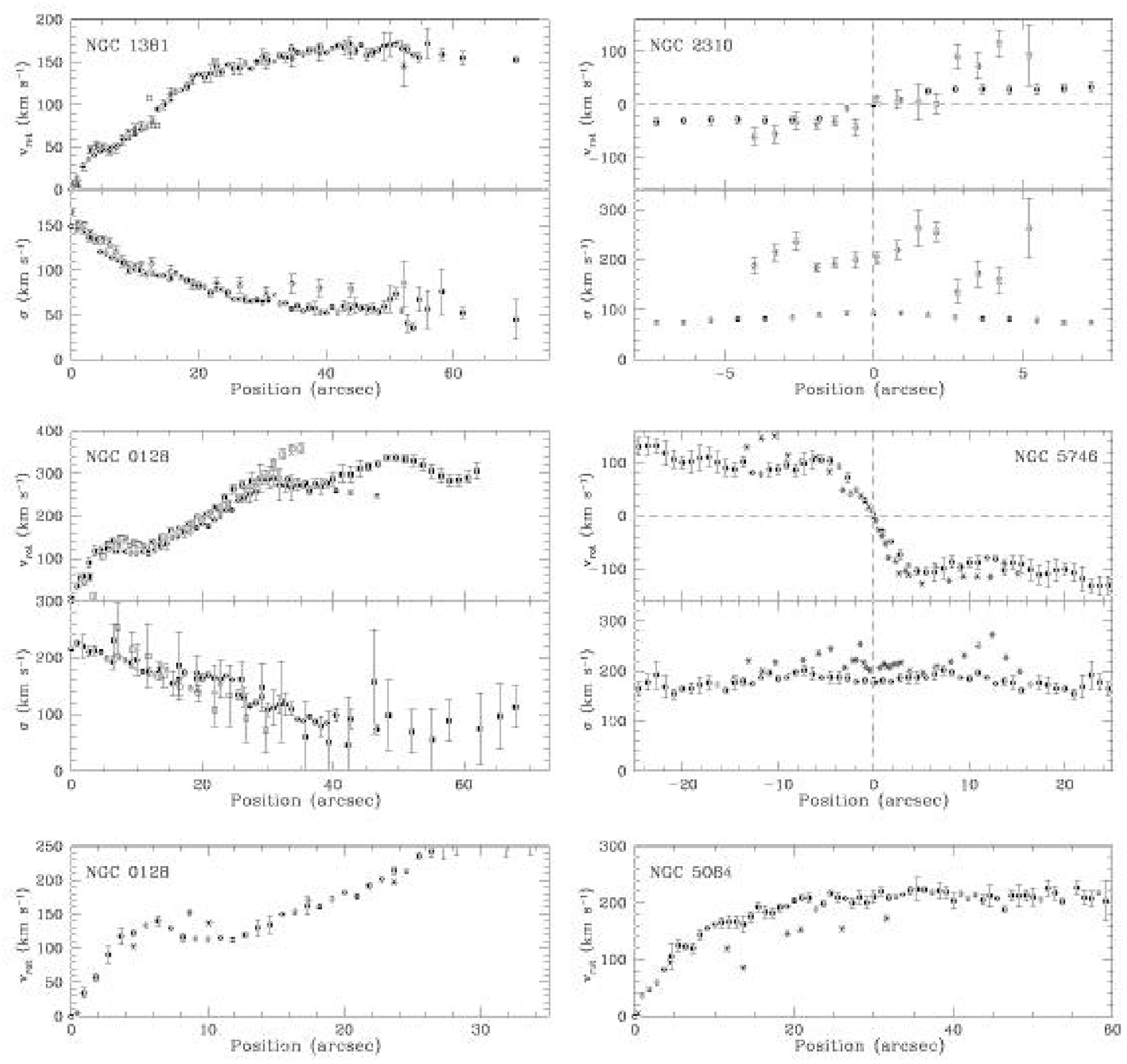}
\epsscale{1.0}
\end{figure}
\clearpage
\begin{figure}
\epsscale{1.0}
\plotone{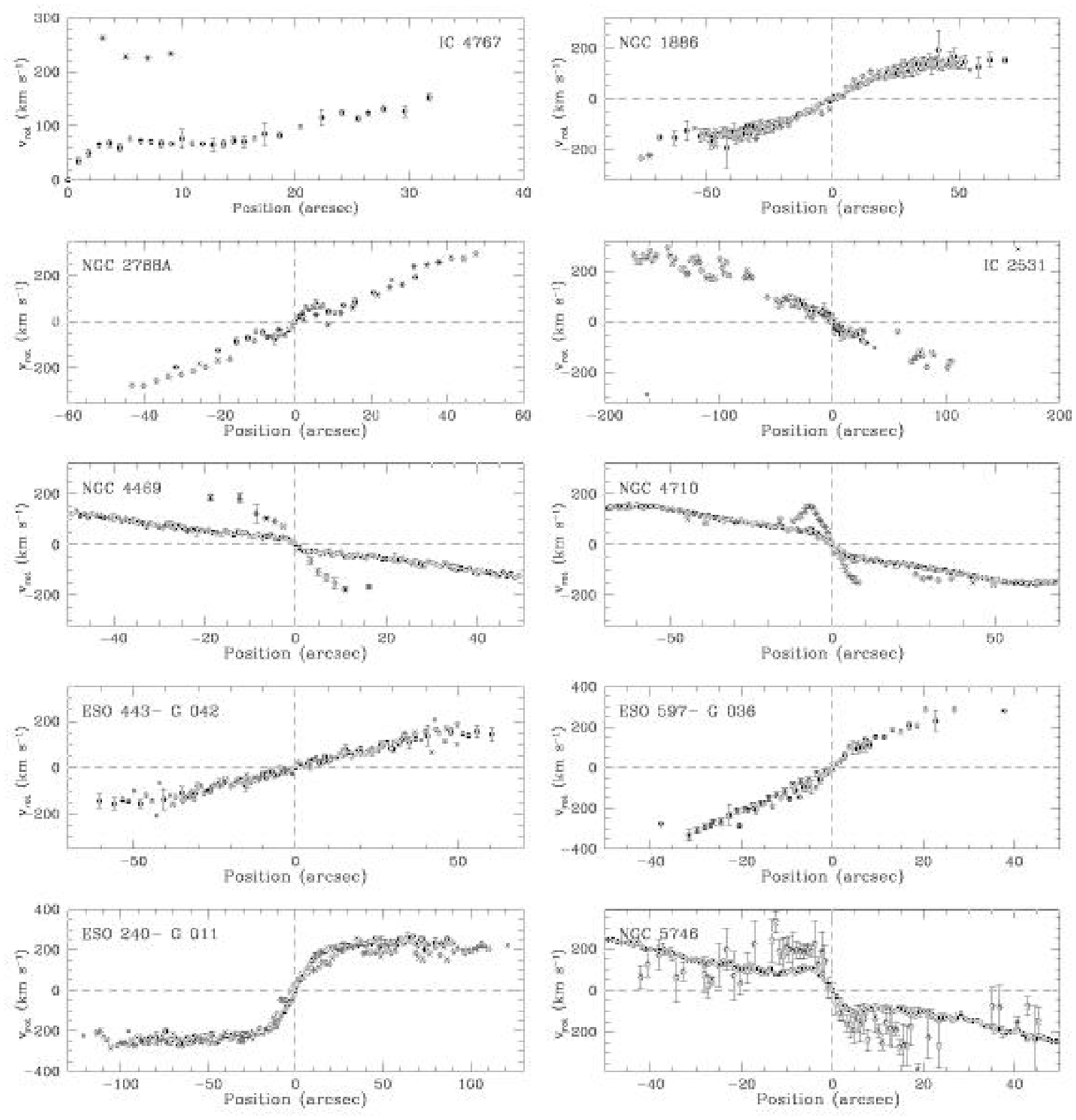}
\epsscale{1.0}
\end{figure}
\clearpage
\begin{figure}
\epsscale{1.0}
\plotone{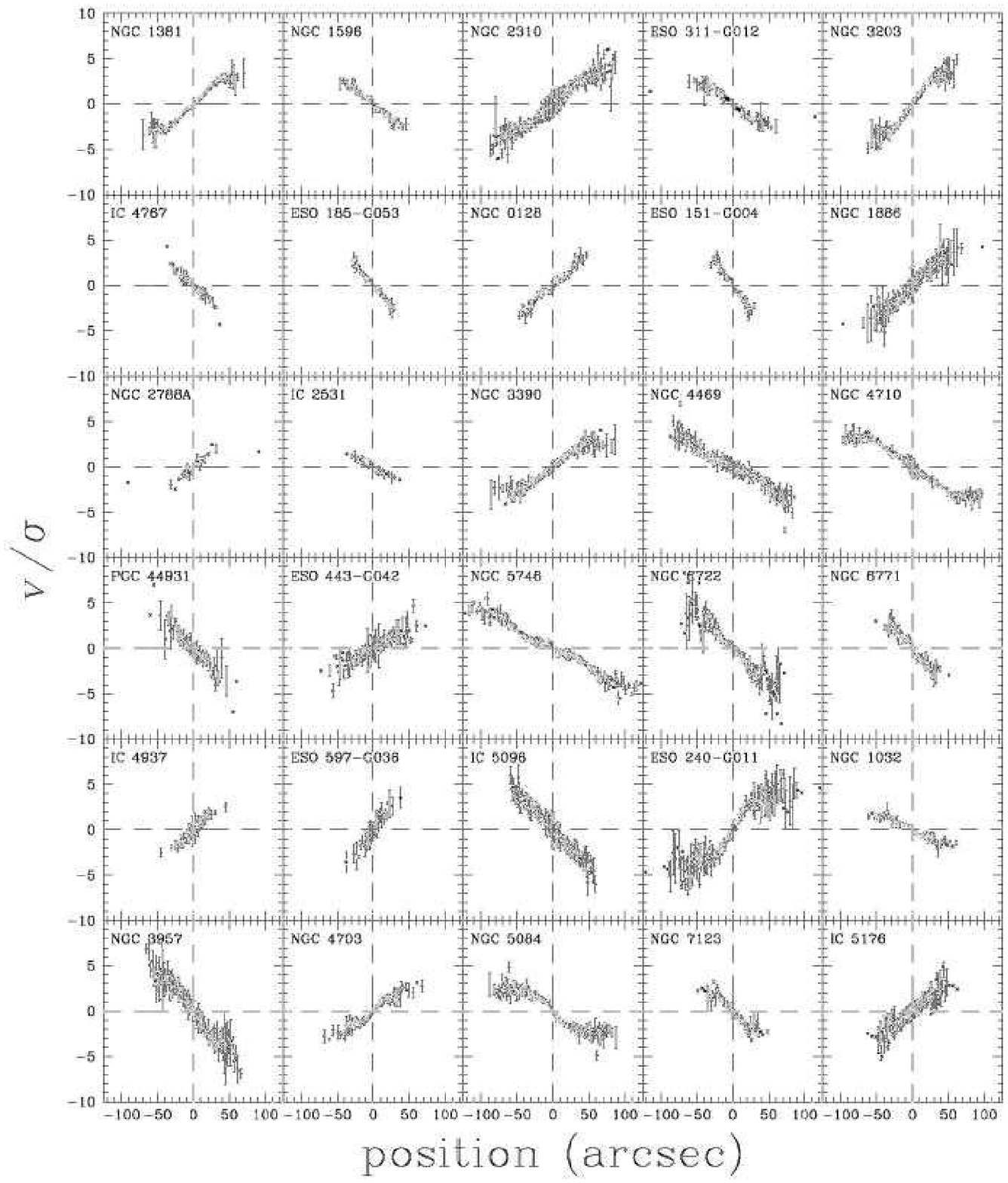}
\epsscale{1.0}
\end{figure}
\clearpage
\begin{figure}
\epsscale{1.0}
\plotone{chung.fig8.eps}
\epsscale{1.0}
\end{figure}
\clearpage
\begin{figure}
\epsscale{1.0}
\plotone{chung.fig9.eps}
\epsscale{1.0}
\end{figure}
\end{document}